\documentclass[11pt]{article}

\usepackage[letterpaper,margin=1in,headheight=14pt]{geometry}

\usepackage[T1]{fontenc}
\usepackage[tt=false]{libertine}
\usepackage[varqu,varl]{zi4}
\usepackage{amsmath}
\usepackage[libertine]{newtxmath}

\usepackage{graphicx}
\usepackage{xcolor}
\usepackage{booktabs}
\usepackage{listings}
\usepackage{multirow}
\usepackage{enumitem}
\usepackage[numbers,sort&compress]{natbib}
\usepackage{upquote}  
\usepackage[hidelinks]{hyperref}
\hypersetup{
  pdftitle={DeQL: A Decision Query Language for Prescriptive Analytics over Relational Data},
  pdfauthor={Matteo Brucato and Fjodor Kholodkov and Soren Little and Jakob Mayer and Duc Nguyen}}
\usepackage{doi}  
\usepackage[capitalise,noabbrev]{cleveref}
\usepackage{caption}
\usepackage{fancyhdr}
\setlist{leftmargin=*, nosep, topsep=2pt}

\newcommand{\systemname}{Decisionhouse}
\newcommand{\querylang}{DeQL}

\newcommand{\kw}[1]{\texttt{#1}}

\newcommand{\kwPREDICT}{\kw{PREDICT}}
\newcommand{\kwDECIDE}{\kw{DECIDE}}

\newcommand{\kwCREATECANDIDATES}{\kw{CREATE CANDIDATES}}

\newcommand{\kwCREATEABSTRACTION}{\kw{CREATE ABSTRACTION}}
\newcommand{\kwDECISIONKEY}{\kw{DECISION KEY}}
\newcommand{\kwDECISIONCOLUMNS}{\kw{DECISION COLUMNS}}
\newcommand{\kwBY}{\kw{BY}}
\newcommand{\kwFILTER}{\kw{FILTER}}

\definecolor{kwcolor}{RGB}{0,80,160}      
\definecolor{strcolor}{RGB}{163,21,21}    
\definecolor{numcolor}{RGB}{128,0,128}    
\definecolor{colcolor}{RGB}{70,70,110}    
\definecolor{relcolor}{RGB}{0,110,70}     
\definecolor{concolor}{RGB}{140,80,140}   
\definecolor{catalogcolor}{RGB}{180,80,0} 
\definecolor{deccolor}{RGB}{180,85,0}     

\newcommand{\colid}[1]{\textcolor{colcolor}{\texttt{#1}}}   
\newcommand{\varid}[1]{\textcolor{deccolor}{\texttt{#1}}}   
\newcommand{\relid}[1]{\textcolor{relcolor}{\texttt{#1}}}   

\lstdefinelanguage{DeQL}[]{SQL}{
  morekeywords=[2]{CANDIDATES, ABSTRACTION, DECIDE, SELECTION,
                   KEY, DECISION, INTO, SUBJECT, TO, CONSTRAINT, MINIMIZE, MAXIMIZE,
                   CONTINUOUS, BINARY, USING, ON, INTEGER, WITH, COLUMNS,
                   FILTER, DISJUNCTIVE, CUMULATIVE, OVERLAPS, TIMESTAMP, INTERVAL,
                   UNBOUNDED, FOREIGN, REFERENCES,
                   PACKAGE, REPEAT, SUCH, THAT,
                   WITHIN, TIMEOUT, STOCHASTIC, PROBABILITY, EXPECTED,
                   PREDICT,
                   LAG, LEAD, FIRST_VALUE, LAST_VALUE, OVER},
  morekeywords=[3]{wh_id, store_id, product_id, route_id, worker_id, shift_id, job_id,
                   item_id, recipe_id, max_servings, sku, region, category, from_wh, to_store, location,
                   cost, demand, price, weight, value, profit, rating, id,
                   capacity, time_hrs, max_qty, max_budget, budget_alloc,
                   skill_level, required_skill, processing_time, deadline, priority, status,
                   machine, kcal, fat, gluten, stock, margin, in_stock, available,
                   perishable, on_sale, channel_id, expected_roi, shipping_cost, warehouse,
                   active_in_last_quarter, employee_id, duration, departure_time, arrival_time,
                   slot_id, slot_date, start_hour, flight_date, holiday_date,
                   batch_id, ready_time,
                   spend_level, expected_return, roi, spend, sector,
                   dividend_yield, earnings_per_share, facility_id, customer_id,
                   open_cost, transport_cost,
                   one_source, nutrient, min_amount, max_amount, protein, fiber,
                   unit_cost, min_nutrient, max_nutrient, num_units,
                   production_cost, min_production, max_production,
                   production_bounds, total_cost, fully_invested,
                   min_holdings, max_holdings, cap_health, min_income,
                   pool_id, workload_id, latency_ms, demand_mean, demand_std,
                   memory_gb, qos_class, mem_per_unit, min_memory_gb,
                   product, period, max_rate, warehouse_cap, max_ramp,
                   opening_stock, safety_stock, hold_cost, run_cost,
                   value_per_hour, max_hours, mismatch},
  morekeywords=[4]{warehouses, stores, routes, products, items, orders, forecasts, shipping,
                   StockItems, Recipes, Products, Workers, Shifts, Catalog, ProductionJobs,
                   MarketingChannels, Crew, Flights, Holidays, TimeSlots,
                   w, r, s, p, f, i, I, R, P,
                   shipping_routes, optimal_shipments, active_routes, shipping_plan, replenishment_items, restock_plan,
                   gluten_free_recipes, meal_plan, worker_shifts, schedule, jobs, production_plan,
                   product_clusters, route_clusters, cart, order, selection, result, potential_routes, channels,
                   spending_plan, plan, training_tasks, training_schedule, training_options,
                   item_batch_assignments, batch_timing, batches, a, t,
                   foods, diet_options, diet_plan, facilities, facility_options, customers, facility_assignments,
                   location_plan, companies, portfolio, production_options, production_plan,
                   gpu_assignments, gpu_pools, workloads, assignments, assignment_options, allocation_plan, spend_options,
                   SpendLevels,
                   timing, Forecast, hourly_demand, capacity_plan, sales, restock, restock_candidates,
                   second_stage, refined_plan, assignment_clusters, TransportCosts, b, c, l, store_demand},
  morekeywords=[5]{supply_limit, meet_demand, time_limit, total_budget, per_channel, min_active,
                   budget, one_shift, one_worker, meet_deadline, resource_cap, no_overlap, count, calories,
                   total_weight, per_category, perishable_weight, no_flight_conflict, no_sunday,
                   no_holiday, no_conflict, one_per_employee, link,
                   min_protein, max_protein, min_fiber,
                   batch_cap, cap_tech, facility_cap, item_deadline, ready,
                   one_batch, pool_capacity, mem_capacity, sla_reserve, sequence, serve, one_level,
                   balance, ramp, opening, closing, cluster_limit},
  morekeywords=[6]{chosen, selected, active, quantity, qty, reserved, spot,
                   assigned, start_time, batch_load, servings, units_produced,
                   opened, amount, is_selected, produce, inventory, hours},
  morekeywords={BY, TRUE, FALSE, SET},
  deletekeywords={HOUR, ZONE, LOCAL, VALUE},
  sensitive=false,
}

\lstdefinestyle{deql}{
  language=DeQL,
  basicstyle=\ttfamily\small\linespread{0.9}\selectfont,
  keywordstyle=\color{kwcolor}\bfseries,          
  keywordstyle=[2]\color{kwcolor}\bfseries,       
  keywordstyle=[3]\color{colcolor},               
  keywordstyle=[4]\color{relcolor},               
  keywordstyle=[5]\color{concolor},               
  keywordstyle=[6]\color{deccolor},               
  identifierstyle=\color{colcolor},               
  commentstyle=\color{gray},
  stringstyle=\color{strcolor},                   
  numberstyle=\color{numcolor},                   
  literate=*{0}{{{\color{numcolor}0}}}{1}         
            {1}{{{\color{numcolor}1}}}{1}
            {2}{{{\color{numcolor}2}}}{1}
            {3}{{{\color{numcolor}3}}}{1}
            {4}{{{\color{numcolor}4}}}{1}
            {5}{{{\color{numcolor}5}}}{1}
            {6}{{{\color{numcolor}6}}}{1}
            {7}{{{\color{numcolor}7}}}{1}
            {8}{{{\color{numcolor}8}}}{1}
            {9}{{{\color{numcolor}9}}}{1},
  breaklines=true,
  showstringspaces=false,
  columns=fullflexible,
  keepspaces=true,
  basewidth={0.5em,0.4em},
  xleftmargin=3pt,
  xrightmargin=3pt,
  frame=single,
  frameround=tttt,
  framerule=0.4pt,
  rulecolor=\color{kwcolor!20},
  backgroundcolor=\color{kwcolor!3},
  aboveskip=0pt,
  belowskip=0pt,
  tabsize=2,
}

\usepackage{etoolbox}
\BeforeBeginEnvironment{lstlisting}{\par\addvspace{\smallskipamount}\noindent\begin{minipage}[t]{\linewidth}}
\AfterEndEnvironment{lstlisting}{\end{minipage}\par\addvspace{\smallskipamount}}

\begin{document}

\title{DeQL: A Decision Query Language for\\ Prescriptive Analytics over Relational Data}

\author{%
  Matteo Brucato \quad Fjodor Kholodkov \quad Soren Little\\
  Jakob Mayer \quad Duc Nguyen\\[4pt]
  OSM Data\\
  \texttt{\{matteo, fjodor, soren, jakob, duc\}@osm-data.com}%
}

\date{Language Specification\\
  Version 0.2, August 2026\\[4pt]
  {\small\textcopyright\ 2026 OSM Data. All rights reserved.}}

\maketitle

\begin{abstract}
\noindent \querylang{} (Decision Query Language) extends SQL to express decision queries: given options drawn from relational data, constraints from policy, and a measurable objective, a \querylang{} query computes the best course of action. Two constructs carry the core of the extension: \kwCREATECANDIDATES{}, which defines the space of options from relational sources, and \kwDECIDE{}, which declares decision variables, named constraints, and an objective over them. The design follows SQL's principles: the user states what to optimize while the engine chooses how to solve it, every query consumes and produces relations, and the structure of a problem stays visible to the engine. This document specifies the language: its design principles, syntax, formal grammar, and execution model. The examples span subset selection, allocation, assignment, scheduling, and decisions at multiple levels of aggregation. The specification also covers extensions for optimization under uncertainty, inline model scoring, and time- and quality-bounded solving. It is an early version of the specification; the language is under active development, and this version fixes the core constructs on which later revisions will build.
\end{abstract}

\tableofcontents
\clearpage

\section{Introduction}
\label{sec:introduction}

Data systems today answer two broad kinds of question. \emph{Descriptive} queries report what the data records: SQL joins and aggregates over warehouses and lakehouses answer ``what happened?'' \emph{Predictive} queries, which can be expressed inside the database through in-database machine learning, estimate ``what will happen?'' A third kind, \emph{prescriptive}, asks what to \emph{do}: which routes to activate, how much capacity to allocate to each workload, which items to select under a budget, when to schedule each task. These are \emph{decision queries}: given options drawn from relational data, constraints from policy, and a measurable objective, a decision query computes the best course of action. SQL has no construct for them; expressing one means dropping to a separate solver or modeling tool and encoding the formulation by hand. \querylang{} (Decision Query Language, pronounced ``dee-kwl'') extends SQL to express decision queries declaratively.

The design rests on four principles inherited from SQL.
\begin{itemize}
\item \textbf{Declarative separation.} The user states \emph{what} to optimize (the variables, constraints, and objective) and the engine determines \emph{how} to solve it, choosing a mathematical formulation and an algorithm (\Cref{sec:pipeline}).
\item \textbf{Relational closure.} A decision query consumes relations and produces a relation. Decisions are data, so results compose with SQL and with further decision queries, as \kw{CREATE TABLE AS} composes in SQL.
\item \textbf{Preserved structure.} \querylang{} keeps a decision problem's structure visible to the engine: which entities a decision concerns, how decisions are grouped, and which decisions each constraint involves stay explicit in the query. The engine reads this structure to recognize problems it can solve with specialized, efficient algorithms (\Cref{sec:pipeline}).
\item \textbf{Conservative extension.} \querylang{} reuses SQL's keywords and semantics wherever they apply, so a reader fluent in SQL meets few new concepts and existing SQL operations behave unchanged.
\end{itemize}

These principles distinguish \querylang{} from modeling languages such as AMPL~\cite{ampl}, Pyomo~\cite{pyomo}, and JuMP~\cite{jump}, which express a chosen formulation over parameters extracted into a separate modeling environment. \querylang{} operates on relational data directly and offers a small set of composable primitives (typed variables, grouped and filtered constraints, and joins that relate decisions made at different levels of aggregation; \Cref{sec:decide-query,sec:constraints,sec:mg-decisions}) that users combine, rather than a fixed catalog of problem templates. It builds on PaQL~\cite{paql}, which introduced declarative package queries, and SolveDB~\cite{solvedb}, which integrated solvers into SQL; \querylang{} generalizes package selection to typed decision variables and grouped constraints and leaves the choice of formulation to the engine.

This document specifies the \querylang{} language; the \systemname{} vision paper~\cite{deql-vision-paper} presents the system that executes it (its architecture, compilation pipeline, and research agenda), and the \querylang{} Studio demonstration~\cite{deql-studio-demo} shows it in use. The remainder specifies the language and its example queries (\Cref{sec:language} onward), the formal grammar (\Cref{sec:grammar}), the execution model (\Cref{sec:pipeline}), and the current scope and limitations (\Cref{sec:discussion}). This is an early version of the specification: it fixes the core constructs, and we expect the language to grow as the system and its applications mature.

\section{The \querylang{} Language}
\label{sec:language}

This section introduces \querylang{} with two short decision queries; the sections that follow specify its constructs in turn, and \Cref{sec:query-examples} collects example queries.

\label{sec:at-a-glance}

Making a decision in \querylang{} involves two statements: \kwCREATECANDIDATES{} defines the options, and a \kwDECIDE{} query chooses among them. The smallest example allocates a fixed pool of GPU-hours across competing workloads:
\begin{lstlisting}[style=deql]
CREATE CANDIDATES workloads
DECISION KEY (workload_id) AS
  SELECT workload_id, value_per_hour, max_hours FROM jobs;

DECIDE plan
FROM workloads
DECISION COLUMNS (hours CONTINUOUS BETWEEN 0 AND max_hours)
SUBJECT TO
  CONSTRAINT cluster_limit: SUM(hours) <= 5000
MAXIMIZE SUM(value_per_hour * hours);
\end{lstlisting}
Each row of \texttt{workloads} is a \emph{candidate}, one option the decision ranges over. \kwDECIDE{} adds a \emph{decision column}, \texttt{hours}, whose value the engine computes for each candidate: how many GPU-hours to assign that workload, any amount between $0$ and its \texttt{max\_hours}. The cluster has a fixed number of hours, so the \texttt{cluster\_limit} constraint caps their total; the objective maximizes the value those hours produce. The result \texttt{plan} is an ordinary relation, one row per workload with its allocated \texttt{hours}: queryable, joinable, and composable with SQL like any table. \Cref{tab:plan-io} runs it on a small input.

\begin{table}[t]
\centering
\footnotesize
\renewcommand{\arraystretch}{1.15}
\caption{The allocation query on a small input. The engine fills the highest-\texttt{value\_per\_hour} workloads to their \texttt{max\_hours} until the 5000-hour capacity is spent, leaving W3 partly allocated and W4 at zero. Every workload appears in \texttt{plan}, including the one at $0$.}
\label{tab:plan-io}
\begin{tabular}[t]{@{}l@{}}
\relid{workloads}~\textit{(input)} \\[3pt]
\begin{tabular}{@{}lrr@{}}
\toprule
\colid{workload\_id} & \colid{value\_per\_hour} & \colid{max\_hours} \\
\midrule
W1 & 9 & 2000 \\
W2 & 7 & 1500 \\
W3 & 5 & 2500 \\
W4 & 3 & 1000 \\
\bottomrule
\end{tabular}
\end{tabular}
\qquad$\rightarrow$\qquad
\begin{tabular}[t]{@{}l@{}}
\relid{plan}~\textit{(result)} \\[3pt]
\begin{tabular}{@{}lr@{}}
\toprule
\colid{workload\_id} & \varid{hours} \\
\midrule
W1 & 2000 \\
W2 & 1500 \\
W3 & 1500 \\
W4 & 0 \\
\bottomrule
\end{tabular}
\end{tabular}
\end{table}

A decision can instead be keep-or-drop. The \kw{SELECTION} modifier marks a decision column as a choice to include or exclude each candidate, as in selecting items under a budget:
\begin{lstlisting}[style=deql]
CREATE CANDIDATES products
DECISION KEY (product_id) AS
  SELECT product_id, price, rating FROM Catalog;

DECIDE cart
FROM products
DECISION COLUMNS (chosen SELECTION BINARY)
SUBJECT TO
  CONSTRAINT budget: SUM(price) <= 1000
MAXIMIZE SUM(rating);
\end{lstlisting}
The constraint bounds total price and the objective maximizes total rating, mirroring the allocation above. The results differ in shape. \texttt{plan} reported a value for every workload, including the one at $0$, because a plain decision column produces one output per candidate. \texttt{cart} instead lists only the chosen products (\Cref{tab:cart-io}): a \kw{SELECTION} variable that is $0$ drops its candidate from the result (\Cref{sec:package-semantics}). This is how \querylang{} expresses package selection (\Cref{sec:paql-compat}).

\begin{table}[t]
\centering
\footnotesize
\renewcommand{\arraystretch}{1.15}
\caption{The selection query on a small input, with the budget at $1000$. The optimum keeps \{P1, P4\} (price $900$, rating $13$); P2 and P3 score lower or exceed the budget, so they are omitted from \texttt{cart}.}
\label{tab:cart-io}
\begin{tabular}[t]{@{}l@{}}
\relid{products}~\textit{(input)} \\[3pt]
\begin{tabular}{@{}lrr@{}}
\toprule
\colid{product\_id} & \colid{price} & \colid{rating} \\
\midrule
P1 & 600 & 9 \\
P2 & 500 & 7 \\
P3 & 450 & 5 \\
P4 & 300 & 4 \\
\bottomrule
\end{tabular}
\end{tabular}
\qquad$\rightarrow$\qquad
\begin{tabular}[t]{@{}l@{}}
\relid{cart}~\textit{(result)} \\[3pt]
\begin{tabular}{@{}lrr@{}}
\toprule
\colid{product\_id} & \colid{price} & \colid{rating} \\
\midrule
P1 & 600 & 9 \\
P4 & 300 & 4 \\
\bottomrule
\end{tabular}
\end{tabular}
\end{table}

The allocation and the cart are deliberately simple first examples. The sections that follow develop the full language: grouped and filtered constraints, decision-column grain, multi-granularity joins, \kw{SELECTION} desugaring, and extensions for uncertainty, inline scoring, and bounded-time solving, which together cover a wide range of decision problems.

\section{Defining Candidates}
\label{sec:defining-candidates}

\subsection{CREATE CANDIDATES}
\label{sec:create-candidates}

\kwCREATECANDIDATES{} defines the decision domain from relational sources. Like a view, it is defined by a query rather than populated directly; like a table, it is a named relation that \kwDECIDE{} queries reference, which is why the text calls it a candidate set when defining it and a candidate table when a query reads it. Beyond both, it carries decision-specific structure: a \kwDECISIONKEY{}, and optionally decision columns and constraints.

\begin{lstlisting}[style=deql]
CREATE CANDIDATES <name>
DECISION KEY (<column_list>) AS
  <select_query>;
\end{lstlisting}

The \kwDECISIONKEY{} specifies which columns identify a single decision entity; the system instantiates one set of decision variables per distinct value. Just as a \kw{PRIMARY KEY} fixes row identity for storage, the \kwDECISIONKEY{} fixes decision identity for optimization, and like a \kw{PRIMARY KEY} its columns must genuinely identify a row: non-\kw{NULL} and unique, so the defining query yields at most one row per value and functionally determines every other column. A candidate that violates this is rejected at materialization with the offending values reported; the defining query must collapse duplicates itself (e.g., with \kw{DISTINCT} or \kw{GROUP BY}).

\noindent\textbf{Example.} Shipping routes as candidates:
\begin{lstlisting}[style=deql]
CREATE CANDIDATES shipping_routes
DECISION KEY (wh_id, store_id) AS
  SELECT r.from_wh AS wh_id, r.to_store AS store_id,
         w.capacity, s.demand, r.cost, r.time_hrs
  FROM routes r
  JOIN warehouses w ON r.from_wh = w.wh_id
  JOIN stores s ON r.to_store = s.store_id;
\end{lstlisting}

The candidate definition accepts any SQL query supported by the underlying system. A candidate set is a first-class object the system owns: it chooses the physical representation (virtual, in-memory, or persisted) by query complexity and reuse, and the structures derived from the candidates, such as the clustering abstractions used for decomposition (\Cref{sec:abstraction}), are attached to it and shared across the queries that reference it.

A flat join of warehouses, stores, and routes does not say whether a decision is made per warehouse, per store, or per route; \texttt{DECISION KEY (wh\_id, store\_id)} fixes it at one decision per warehouse-store pair, and from there governs how many variables are created, how constraints group, and how results map back to tuples.

A per-column grain (the \kwBY{} clause of \Cref{sec:variable-grain}) sets how coarsely a single column is decided. The \kwDECISIONKEY{}, in contrast, grains the entire candidate set, and it gives \querylang{} three properties a per-column grain alone does not provide. First, data columns derive their grain from it: in \texttt{shipping\_routes}, \texttt{cost} is single-valued per warehouse-store pair, so a per-row constraint over it instantiates once per pair. Second, the key is an integrity constraint: \texttt{DECISION KEY (wh\_id, store\_id)} asserts one decision per warehouse-store pair, and a defining query that produces two rows for a pair is rejected rather than silently double-counted. Third, it fixes the result's shape, one row per key value (or the chosen subset, for a \kw{SELECTION} decision), and is what a \kw{FOREIGN DECISION KEY} references when decisions compose across candidate sets (\Cref{sec:multi-granularity}). Modeling languages index each variable family on a set of its own, as in \texttt{var x\{Sites, Hours\}} in AMPL; in \querylang{} every index set a query names is some candidate table's \kwDECISIONKEY{}, coarsened at most by \kwBY{}.

Where \kwDECISIONKEY{} says what each decision is \emph{about}, \kwDECISIONCOLUMNS{} declares what is \emph{decided} for it. An optional \kwDECISIONCOLUMNS{} clause can be included directly in \kwCREATECANDIDATES{} to declare default decision columns at the schema level:

\begin{lstlisting}[style=deql]
CREATE CANDIDATES shipping_routes
DECISION KEY (wh_id, store_id)
DECISION COLUMNS (
    active   SELECTION BINARY,
    quantity CONTINUOUS BETWEEN 0 AND capacity)
AS SELECT ...;
\end{lstlisting}

When \kwDECISIONCOLUMNS{} is declared at the schema level, \kwDECIDE{} queries can omit variable declarations and inherit them from the candidate definition.

\subsection{Schema-Level Constructs}
\label{sec:schema-level}

\kwCREATECANDIDATES{} can include schema-level \kwDECISIONCOLUMNS{} and \kw{CONSTRAINTS} blocks that are automatically enforced in any \kwDECIDE{} query referencing the candidates.

\begin{lstlisting}[style=deql]
CREATE CANDIDATES item_batch_assignments
DECISION KEY (item_id, batch_id)
FOREIGN DECISION KEY (batch_id)
  REFERENCES batch_timing(batch_id)
DECISION COLUMNS (
  assigned SELECTION BINARY,
  batch_load = SUM(weight * assigned)
    REFERENCES batch_timing(batch_id))
CONSTRAINTS (
  one_batch: SUM(assigned) = 1 BY item_id,
  batch_cap: batch_load <= capacity)
AS SELECT ...;
\end{lstlisting}

\begin{itemize}
\item \textbf{\kw{FOREIGN DECISION KEY}} declares that some of this candidate set's key columns reference another candidate set's \kwDECISIONKEY{}, naming the coarser level that a derived column or a joined \kwDECIDE{} can reach (\Cref{sec:mg-decisions}). The optional \kw{AS} alias names that level, so declarations in the same statement can qualify its columns when a name is ambiguous.
\item \textbf{Plain decision columns} are declared exactly as in a \kwDECIDE{} query, as in \texttt{assigned SELECTION BINARY}.
\item \textbf{Derived decision columns} use \texttt{=} syntax to define a variable whose value is determined by an aggregate over related rows. For example, \texttt{batch\_load = SUM(weight * assigned)} with \texttt{REFERENCES batch\_timing(batch\_id)} creates a batch-level variable equal to the sum of assigned item weights. The \kw{REFERENCES} clause specifies which foreign candidate table the derived variable belongs to.
\item \textbf{\kw{CONSTRAINTS}} declares structural constraints that are always enforced. Schema-level constraints are immutable: they cannot be overridden or relaxed in \kwDECIDE{} queries. Structural constraints that always hold are declared once in the schema, while problem-specific business constraints go in \kwDECIDE{} queries. For example, a workforce scheduling schema might enforce that no employee works more than one shift at a time, a physical impossibility rather than a business preference.
\end{itemize}

\subsection{Candidate Abstractions}
\label{sec:abstraction}

\kwCREATEABSTRACTION{} defines a coarser view of a candidate set by grouping candidates into representative clusters. An abstraction lets the engine solve a reduced problem over the representatives and then refine within promising groups, trading bounded optimality loss for scalability on large candidate sets (\Cref{sec:pipeline}).

\begin{lstlisting}[style=deql]
CREATE ABSTRACTION assignment_clusters
ON gpu_assignments (cost, demand, capacity)
USING KMEANS(k=1000);
\end{lstlisting}

The \kw{ON} clause names the candidate set and the columns used for clustering; the \kw{USING} clause names the clustering method and its parameters ($k$-means into 1000 groups here). An abstraction is a derived object, much like a database index: it does not alter the candidate set, and \kwDECIDE{} queries continue to reference the original candidates. Where an index returns the same answer faster, an abstraction returns a faster approximate one, at the bounded optimality loss above.

\section{The DECIDE Query}
\label{sec:decide-query}

\subsection{The DECIDE Statement}
\label{sec:decide-statement}

The \kwDECIDE{} statement is the core prescriptive query. It optimizes over a candidate set and returns a result relation. The result extends the candidate columns with the declared decision columns, populated with the optimal assignment. Its anatomy:

\begin{lstlisting}[style=deql]
DECIDE <result_name>
FROM <candidate_ref> [JOIN ...]
[DECISION COLUMNS (<variable_declarations>)]
[WHERE <filter_predicate>]
SUBJECT TO
  <constraint_list>
[MINIMIZE | MAXIMIZE <objective_expression>];
\end{lstlisting}

\begin{itemize}
\item \textbf{\kw{FROM}} references one or more candidate sets, each either a named set created by \kwCREATECANDIDATES{} or a table or subquery given an inline \kwDECISIONKEY{}.
\item \textbf{\kwDECISIONCOLUMNS{}} declares \emph{decision columns}: new columns of the result, one per candidate entity, whose values the engine computes (the \emph{decision variables}). Each declares a name, type (\kw{BINARY}, \kw{INTEGER}, \kw{CONTINUOUS}), optional \kw{SELECTION} modifier, optional bounds (\kw{BETWEEN} $\ell$ \kw{AND} $u$), and an optional \kwBY{} grain (\Cref{sec:variable-grain}).
\item \textbf{\kw{WHERE}} filters candidates before optimization (predicate pushdown: candidates not satisfying the predicate are excluded from the decision domain). The predicate may reference only data columns; decision columns have no values until the query is solved.
\item \textbf{\kw{SUBJECT TO}} introduces the constraint block: a comma-separated list of constraints, each optionally named.
\item \textbf{\kw{MINIMIZE}/\kw{MAXIMIZE}} specifies the objective function as an aggregate expression over candidate columns and decision variables. The objective must have the empty grain, i.e., denote a single value: an ungrouped aggregate, or terms built from \texttt{BY ()} variables and constants (\Cref{sec:variable-grain}). The objective is optional: a \kwDECIDE{} with no \kw{MINIMIZE} or \kw{MAXIMIZE} is a feasibility query, returning any solution that satisfies the constraints. \kw{MINIMIZE}~\kw{MAX} states a minimax objective and \kw{MAXIMIZE}~\kw{MIN} a maximin one; both are realized by the standard reformulation, an auxiliary variable that replaces the aggregate, bounded by one inequality per distinct value of the aggregated expression's grain (\Cref{sec:variable-grain}). The reverse pairings, \kw{MINIMIZE}~\kw{MIN} and \kw{MAXIMIZE}~\kw{MAX}, are not admitted in this version: no such reformulation is exact for them, since forcing the auxiliary onto one term needs a binary variable per term.
\end{itemize}

\noindent\textbf{Example.} Shipping plan with mixed variables:
\begin{lstlisting}[style=deql]
DECIDE shipping_plan
FROM shipping_routes
DECISION COLUMNS (
    active   SELECTION BINARY,
    quantity CONTINUOUS BETWEEN 0 AND capacity)
WHERE time_hrs <= 48
SUBJECT TO
  CONSTRAINT supply_limit: SUM(quantity) <= capacity BY wh_id,
  CONSTRAINT meet_demand:  SUM(quantity) = demand BY store_id
MINIMIZE SUM(cost * quantity);
\end{lstlisting}

The \kw{SELECTION BINARY} modifier on \texttt{active} provides auto-multiply and interval-linking semantics (\Cref{sec:package-semantics}).

\noindent\textbf{Inline candidate sets.} A candidate set named with \kwCREATECANDIDATES{} can be reused across queries, can carry schema-level constraints enforced on every decision over it (\Cref{sec:schema-level}), and can be referenced by a \kw{FOREIGN DECISION KEY} (\Cref{sec:multi-granularity}). When a decision is one-off and needs none of these, a \kwDECIDE{} query may define its options inline, attaching the \kwDECISIONKEY{} to a table or subquery, as PaQL~\cite{paql} states the options and the decision in one query:
\begin{lstlisting}[style=deql]
DECIDE cart
FROM (SELECT product_id, price, rating FROM Catalog)
DECISION KEY (product_id)
DECISION COLUMNS (chosen SELECTION BINARY)
SUBJECT TO CONSTRAINT budget: SUM(price) <= 1000
MAXIMIZE SUM(rating);
\end{lstlisting}
Defining the candidate set by name parallels a view; inlining it parallels a subquery.

\subsection{Decision Column Grain}
\label{sec:variable-grain}

A decision column's \emph{grain} is the set of columns that determine how many variables it expands into: one per distinct combination of their values. By default the grain is the candidate's \kwDECISIONKEY{}, giving one variable per row. A \kwBY{} clause instead groups the column like \kw{GROUP BY}, one variable per group: \texttt{BY region} gives one variable per region, \texttt{BY (region, zone)} one per region-zone combination, and \texttt{BY ()} a single variable over the whole set. Because grain columns are data columns fixed by the \kwDECISIONKEY{}, a \kwBY{} grain is always coarser than the per-row default, never finer. The same \kwBY{} forms apply to grouped constraints (\Cref{sec:constraint-syntax}). How many variables exist depends on the data: the groups are the distinct combinations that survive the \kwDECIDE{} query's \kw{WHERE} filter and, when the query has a join, that survive the join as well (\Cref{sec:multi-granularity}); a group with no surviving rows gets no variable, and a \kw{NULL} in a grain column is an error (\Cref{sec:validity}).

Each column's grain reads off its \kwBY{} clause; for a candidate set keyed on \texttt{(region, sector, store)}:
\begin{lstlisting}[style=deql]
DECISION COLUMNS (
    reserved CONTINUOUS BY (),                -- one variable total
    quantity CONTINUOUS BY region,            -- one per region
    amount   CONTINUOUS BY (region, sector),  -- one per region-sector
    spot     CONTINUOUS)                      -- one per row (the key)
\end{lstlisting}

\noindent\textbf{Grain is per column.} A \kwBY{} grain sets how coarsely a single column is decided; the \kwDECISIONKEY{} still identifies the decision and fixes the result shape at one row per key value (\Cref{sec:create-candidates}). The key spans the whole candidate set, so different decision columns may carry different grains while sharing it.

\noindent\textbf{Example.} Reserve a baseline capacity per region, then meet each store's demand from that reservation or from spot capacity:
\begin{lstlisting}[style=deql]
CREATE CANDIDATES store_demand
DECISION KEY (region, store_id) AS
  SELECT region, store_id, demand, spot_price FROM Stores;

DECIDE plan
FROM store_demand
DECISION COLUMNS (
    reserved CONTINUOUS BY region,
    spot     CONTINUOUS)
SUBJECT TO
  CONSTRAINT meet_demand: reserved + spot >= demand
MINIMIZE 50 * SUM(reserved) + SUM(spot_price * spot);
\end{lstlisting}
\texttt{reserved} is one variable per region (\texttt{BY region}); \texttt{spot} is one per store. The per-row constraint \texttt{reserved + spot >= demand} applies a region's single \texttt{reserved} to each store in it, and \texttt{SUM(reserved)} in the objective contributes one term per region, not one per store. The result has one row per \kwDECISIONKEY{} value, one per store; a coarse-grain variable appears constant within each of its groups, so \texttt{reserved} repeats across a region's stores (\Cref{tab:grain-io}) and is read back with \texttt{SELECT DISTINCT region, reserved FROM plan}. A grain that cross-cuts the key (say, a store tier spanning regions) repeats its value across every store of that tier, coupling stores in different regions.

\begin{table}[t]
\centering
\footnotesize
\renewcommand{\arraystretch}{1.15}
\caption{The regional plan on a small input (\texttt{spot\_price} $= 30$ throughout). \texttt{reserved} is one variable per region: it is constant within each region (West $= 4$, East $= 6$), repeating across that region's stores, while \texttt{spot} tops up each store's demand above the regional baseline.}
\label{tab:grain-io}
\begin{tabular}[t]{@{}l@{}}
\relid{store\_demand}~\textit{(input)} \\[3pt]
\begin{tabular}{@{}llr@{}}
\toprule
\colid{region} & \colid{store\_id} & \colid{demand} \\
\midrule
West & S1 & 10 \\
West & S2 & 4 \\
East & S3 & 8 \\
East & S4 & 6 \\
\bottomrule
\end{tabular}
\end{tabular}
\qquad$\rightarrow$\qquad
\begin{tabular}[t]{@{}l@{}}
\relid{plan}~\textit{(result)} \\[3pt]
\begin{tabular}{@{}llrr@{}}
\toprule
\colid{region} & \colid{store\_id} & \varid{reserved} & \varid{spot} \\
\midrule
West & S1 & 4 & 6 \\
West & S2 & 4 & 0 \\
East & S3 & 6 & 2 \\
East & S4 & 6 & 0 \\
\bottomrule
\end{tabular}
\end{tabular}
\end{table}

\noindent\textbf{Grain of expressions.} Every expression has a grain. A constant has the empty grain \texttt{()}; a data column has the full \kwDECISIONKEY{} grain of the candidate table it comes from, even when its values repeat; a decision column has its declared grain; and a scalar operator's result has the union of its operands' grains, so a coarser operand is broadcast across the finer ones, taking the same value on every row of its group. This applies to decision columns too: in \texttt{var1 * var2} with \texttt{var1} coarser than \texttt{var2}, the single \texttt{var1} is broadcast onto each finer \texttt{var2} row. An aggregate's result has the empty grain, or the grouping columns' grain inside a constraint grouped \kwBY{} (\Cref{sec:constraint-syntax}). Aggregates do not nest, as in SQL.

\noindent\textbf{Aggregation over grain.} An aggregate ranges over the distinct values of its argument's grain, not over the candidate rows. In the objective above, \texttt{SUM(reserved)} with \texttt{reserved BY region} contributes one term per region rather than one per store, because it ranges over \texttt{reserved}'s grain; had \texttt{reserved} been declared \texttt{BY ()}, the same \texttt{SUM(reserved)} would collapse to a single term. This extends the rule that decision variables are not duplicated across rows (\Cref{sec:multi-granularity}) to aggregates: a coarse variable summed over a finer candidate set is counted once per its own group.

\noindent\textbf{Mixed grains.} When the operands have different grains, the union rule above sets what the aggregate ranges over. \texttt{SUM(demand * reserved)}, with \texttt{demand} at the row (store) grain and \texttt{reserved BY region}, ranges over the stores, and each region's single \texttt{reserved} enters with a coefficient equal to the total demand over that region's stores. In general the aggregate runs once over each distinct combination of its argument's grain, reading each operand at that combination. This row-weighted reading is usually the intent: a per-store cost multiplying a regional reservation charges that reservation once for every store it covers. When a charge should instead fall once per group, the cost column must sit on a candidate at that grain, as in the facility-location example (\Cref{sec:example-facility-location}).

\noindent\textbf{Divergences from SQL.} Two consequences follow, and both differ from SQL. An aggregate of a constant collapses to the constant: \texttt{SUM(1)} is 1, since a constant has grain \texttt{()}; use \kw{COUNT(*)} to count candidates. And \kw{AVG} over a coarse argument divides by the number of groups, not the number of rows.
\subsection{SELECTION Semantics}
\label{sec:package-semantics}

The \kw{SELECTION} modifier on a decision column applies two automatic transformations (desugaring rules detailed in \Cref{tab:desugaring}). PaQL~\cite{paql} introduced the first, auto-multiply; \querylang{} adds the second, interval linking.

\noindent\textbf{Auto-multiply.} An additive aggregate over data columns (columns from the candidate definition, not decision variables) is automatically multiplied by the selection variable; the rule covers \kw{SUM} and \kw{COUNT(*)}. With \texttt{active SELECTION BINARY}:
\begin{itemize}
\item \texttt{SUM(cost)} in a constraint becomes \texttt{SUM(cost * active)}
\item \texttt{COUNT(*)} becomes \texttt{SUM(active)}
\end{itemize}
An aggregate whose argument already contains a decision variable, such as \texttt{SUM(cost * quantity)}, is not rewritten: multiplying it by the selection variable would make the expression nonlinear, and for a companion the rewrite would also be redundant, since interval linking already zeroes it when its candidate is unselected. \kw{MIN}, \kw{MAX}, and \kw{AVG} are not rewritten either, since scaling their argument by the selection variable does not give the value over the selected candidates: an unselected candidate would contribute a zero rather than drop out. A bound on the selected candidates' average is written in the linear form \texttt{SUM(kcal * active) <= 500 * SUM(active)}.

\noindent\textbf{Interval linking.} A \emph{companion} is a decision column other than the selection variable, declared at the selection variable's grain on the same candidate table. A companion is active only when its entity is selected: both of its declared bounds scale with the selection variable, so its effective domain is $\{0\} \cup [\ell, u]$. For a companion $x$ declared \kw{BETWEEN} $\ell$ \kw{AND} $u$ within the rule's scope (\textbf{Scope}, below), the system generates $\ell \cdot \texttt{active} \le x \le u \cdot \texttt{active}$ and relaxes the variable's own domain to $[\min(\ell,0), \max(u,0)]$; the declared $[\ell, u]$ is the range \emph{when selected}, and without the relaxation a companion with $\ell > 0$ could never be deselected. For \texttt{quantity CONTINUOUS BETWEEN 0 AND capacity} the rule generates $\texttt{quantity} \le \texttt{capacity} \cdot \texttt{active}$, the lower side being vacuous; for \texttt{weight CONTINUOUS BETWEEN 0.01 AND 0.08} it generates $0.01 \cdot \texttt{active} \le \texttt{weight} \le 0.08 \cdot \texttt{active}$, so an unselected entity forces \texttt{weight} to $0$ and a selected one holds it in $[0.01, 0.08]$. An infinite bound cannot appear in a generated product: the engine links such a companion through an indicator constraint ($\texttt{active} = 0 \Rightarrow x = 0$) or through a finite bound implied by the constraints. An empty band ($\ell > u$) is rejected (\Cref{sec:grammar-types}).

\noindent\textbf{Scope.} Both rules are scoped to one candidate table: a \kwDECIDE{} may declare at most one selection variable per candidate table (\Cref{sec:grammar-types}), and each governs its own table alone. Auto-multiply rewrites only aggregates whose argument comes from the selection variable's table; interval linking reaches only companion columns at the selection variable's own grain on that table. A column with a coarser \kwBY{} grain, or one tagged \kw{ON} another table (\Cref{sec:multi-granularity}), is shared by rows whose selection variables differ; no single selection variable governs it, and forcing it to zero because one of its rows went unselected would be wrong. Tying such a column to a selection is written out as an ordinary constraint, as \texttt{ready} does in \Cref{sec:multi-granularity}.

\noindent\textbf{Bounded repetition.} A \kw{SELECTION INTEGER} variable counts how many copies of a candidate to take, with the range set by ordinary bounds:
\begin{lstlisting}[style=deql]
DECISION COLUMNS (qty SELECTION INTEGER BETWEEN 0 AND 5)
\end{lstlisting}
This lets each candidate appear 0 to 5 times, the integer-multiplicity counterpart of PaQL's \texttt{REPEAT} (\Cref{sec:paql-compat}). Aggregates then count each copy: \kw{COUNT(*)} becomes \texttt{SUM(qty)}, and \texttt{SUM(cost)} becomes \texttt{SUM(cost * qty)}. Interval linking applies to companions of a \kw{SELECTION BINARY} variable only; a repetition count multiplies aggregates but does not link companions (\Cref{sec:grammar-types}).

\noindent\textbf{Result rows.} Beyond rewriting constraints, \kw{SELECTION} shapes the result: a candidate whose selection variable is $0$ is omitted, so a \kwDECIDE{} query with a \kw{SELECTION} column returns the selected subset, one row per chosen candidate, generalizing PaQL's package semantics. A query without a \kw{SELECTION} column keeps one row per candidate, each carrying its computed decision values. The selection variable itself is $1$ on every surviving row, so the result does not carry it as a column; it carries the candidate columns and the remaining decision columns.
When a selection variable's only occurrences after desugaring are the linking constraints of \Cref{tab:desugaring}, distinct optimal solutions can disagree on it: a candidate may be selected with all of its companions at zero. To keep the result relation well-defined, the engine then reports the \emph{minimal} selection, so a row is included exactly when some companion variable is nonzero. The rule reads the companions, and a selection variable that has none cannot reach this case: with no companion to link and no explicit reference it would occur nowhere, and every constraint, and the objective when one is given, must reference a decision variable (\Cref{sec:validity}).

\noindent\textbf{Redundant selection.} The desugaring rules make precise when a selection variable carries no mathematical content. Suppose auto-multiply rewrites no aggregate over its candidate table, every companion's declared band contains zero ($\ell \le 0 \le u$, so a deselected candidate lands inside the companion's own declared domain and the domain relaxation of interval linking leaves it unchanged), and the query never references the selection variable explicitly. Deactivating a candidate is then equivalent to setting its companions to zero, so the program with the selection variable and the program without it admit the same companion assignments and the same optimal value, and the minimal selection above recovers the variable from the companions. The engine may therefore plan and solve such a query without instantiating the selection variable, and structure detection (\Cref{sec:pipeline}) analyzes the companion structure alone. Each hypothesis matters: a band that excludes zero, as with $\ell > 0$, makes the companion semi-continuous, which the language expresses only through the selection variable; an auto-multiplied aggregate, as in the cart query of \Cref{sec:at-a-glance}, makes the selection variable the decision itself; and an explicit reference, such as a fixed activation cost in the objective, prices the selection directly.

\begin{table}[t]
\centering
\footnotesize
\renewcommand{\arraystretch}{1.15}
\caption{\kw{SELECTION} desugaring rules. Given a selection variable $p$ and a companion variable $x$ with bounds $[\ell, u]$, declared at $p$'s grain on $p$'s candidate table, the system automatically generates the constraints shown; columns outside that scope are not rewritten. Auto-multiply covers the additive aggregates \kw{SUM} and \kw{COUNT(*)}; interval linking applies to companions of a \kw{SELECTION BINARY} variable.}
\label{tab:desugaring}
\begin{tabular}{@{}lll@{}}
\toprule
\textbf{Rule} & \textbf{Effect} & \textbf{Generated} \\
\midrule
Auto-multiply & scale an additive data-column aggregate by $p$ & \texttt{SUM(col)} $\to$ \texttt{SUM(col * p)} \\
Interval linking & scale a companion's band by $p$, zero when unselected & $\ell \cdot p \le x \le u \cdot p$ \\
\kw{COUNT(*)} & count selected items, with multiplicity & \texttt{COUNT(*)} $\to$ \texttt{SUM(p)} \\
\bottomrule
\end{tabular}
\end{table}

\subsection{Composability}
\label{sec:composability}

\kwDECIDE{} creates a standard table (analogous to \kw{CREATE TABLE AS}). This enables three forms of composition:

\noindent\textbf{SQL on results.} Decision results can be queried, joined, and aggregated with standard SQL:
\begin{lstlisting}[style=deql]
SELECT wh_id, SUM(quantity) AS total_shipped
FROM shipping_plan
GROUP BY wh_id;
\end{lstlisting}

\noindent\textbf{Decision chaining.} The output of one \kwDECIDE{} becomes input to another, enabling sequential decision-making:
\begin{lstlisting}[style=deql]
CREATE CANDIDATES second_stage
DECISION KEY (wh_id, store_id) AS
  SELECT * FROM shipping_plan WHERE quantity > 0;

DECIDE refined_plan
FROM second_stage ...;
\end{lstlisting}

\noindent\textbf{CTE integration.} A \kwDECIDE{} can be a named step in a \kw{WITH} clause, so SQL steps and a decision step compose in one query. Below, one step forecasts demand in ordinary SQL, a \kwDECIDE{} step plans the restock within a budget, and a final \kw{SELECT} combines the two:
\begin{lstlisting}[style=deql]
WITH forecast AS (
  SELECT product_id, AVG(units_sold) AS demand
  FROM sales GROUP BY product_id
),
restock AS (
  DECIDE FROM restock_candidates
  DECISION COLUMNS (qty INTEGER BETWEEN 0 AND max_qty)
  SUBJECT TO CONSTRAINT budget: SUM(unit_cost * qty) <= 50000
  MAXIMIZE SUM(qty)
)
SELECT f.product_id, f.demand, r.qty
FROM forecast f JOIN restock r ON f.product_id = r.product_id;
\end{lstlisting}
A \kwDECIDE{} written this way takes the step's name as its result table. Each \kwDECIDE{} is solved independently, and its result is an ordinary table to the rest of the query; two decisions in separate steps cannot share a constraint or objective, so coupling them requires a single \kwDECIDE{} over joined candidates (\Cref{sec:multi-granularity}).

The relational closure property (every \querylang{} operator consumes and produces relations) mirrors SQL's composability, so decision queries compose into existing SQL pipelines.
\section{Constraints}
\label{sec:constraints}

\subsection{Constraint Syntax}
\label{sec:constraint-syntax}

Constraints in \querylang{} keep their structure visible: each may be named, grouped with \kwBY{}, and filtered with \kwFILTER{}. A name (\texttt{CONSTRAINT supply\_limit: ...}) is optional; it supports governance (reporting which constraint was violated), debugging (relaxing a constraint and re-solving), and documentation. An unnamed constraint is written as its body alone.

Aggregate constraints, the most common form, follow the pattern:
\begin{lstlisting}[style=deql]
CONSTRAINT <name>: <aggregate_expr> [FILTER (WHERE <predicate>)]
    <comparison_op> <bound_expr>
    [BY <col_list>]
\end{lstlisting}

\noindent\textbf{Aggregate constraints.} An aggregate constraint applies a standard SQL aggregate (\kw{SUM}, \kw{COUNT}, \kw{MIN}, \kw{MAX}, \kw{AVG}) to an expression over candidate columns and decision variables; \kw{MIN} and \kw{MAX} over decision variables are admitted in the direction that expands into one inequality per term, \kw{MAX} under $\leq$ and \kw{MIN} under $\geq$, matching the objective rule (\Cref{sec:decide-statement}):
\begin{lstlisting}[style=deql]
CONSTRAINT budget: SUM(price) <= 100
\end{lstlisting}

\noindent\textbf{Grouped constraints (\kwBY{}).} The \kwBY{} clause on an aggregate constraint generates one constraint per group value, analogous to \kw{GROUP BY}:
\begin{lstlisting}[style=deql]
CONSTRAINT supply_limit: SUM(quantity) <= capacity BY wh_id
\end{lstlisting}
With 10 warehouses, this generates 10 constraints, each bounding the total quantity shipped from one warehouse. The bound is evaluated per group, so every data column it references must be single-valued within each group (here, \texttt{wh\_id} must functionally determine \texttt{capacity}); this is the same single-value rule SQL applies to non-grouped columns under \kw{GROUP BY}, and a constraint whose bound varies within a group is rejected.

\noindent\textbf{Filtered constraints (\kwFILTER{}).} The \kwFILTER{} clause restricts which candidates participate:
\begin{lstlisting}[style=deql]
CONSTRAINT cap_tech: SUM(weight) FILTER (WHERE sector = 'Technology')
  <= 0.25
\end{lstlisting}

\kwBY{} and \kwFILTER{} compose: \texttt{SUM(qty) FILTER (WHERE region = 'US') <= limit BY category} generates one constraint per category, considering only US-region candidates. \kwFILTER{} is distinct from the query-level \kw{WHERE} (\Cref{sec:decide-statement}): \kw{WHERE} removes candidates from the whole problem, whereas \kwFILTER{} scopes a single aggregate, so a candidate it excludes still exists and can participate in other constraints.

\noindent\textbf{Per-row constraints.} Expressions without aggregates generate one constraint instance per distinct value of the expression's grain (\Cref{sec:variable-grain}). A \kwBY{} clause on a per-row constraint does not change how many instances it generates, since the grain already fixes that count. Its only effect is to supply the grouping for any window functions the constraint contains (\Cref{sec:window-constraints}). When the operands are at the full \kwDECISIONKEY{} grain, this is one constraint per candidate row; coarser operands give one per group:
\begin{lstlisting}[style=deql]
CONSTRAINT meet_deadline: start_time + processing_time <= deadline
\end{lstlisting}

\noindent\textbf{Scheduling constraints.} Two constraint forms borrow the global-constraint semantics of constraint programming languages such as MiniZinc~\cite{minizinc}. \kw{DISJUNCTIVE}$(s, d)$ takes a start time $s$ and a duration $d$ and forbids overlap: within each group, the half-open intervals $[s,\, s+d)$ are pairwise disjoint, so no two tasks on a machine run at once:
\begin{lstlisting}[style=deql]
CONSTRAINT no_overlap:
  DISJUNCTIVE(start_time, processing_time) BY machine
\end{lstlisting}
\kw{CUMULATIVE}$(s, d, r)$ adds a per-task resource demand $r$ and bounds concurrent use: within each group, at every instant $t$ the total $r$ over the tasks running then ($s \le t < s+d$) is at most the bound:
\begin{lstlisting}[style=deql]
CONSTRAINT resource_cap:
  CUMULATIVE(start_time, processing_time, demand) <= max_capacity
\end{lstlisting}
A scheduling constraint follows the same grain rule as an aggregate (\Cref{sec:variable-grain}): its tasks are the distinct values of its operands' combined grain, not the candidate rows, so operands carried on a coarser candidate set give one task per group even when a join repeats them across many rows. Each accepts a \kwBY{} clause, as \texttt{no\_overlap} groups by \texttt{machine}; with no \kwBY{}, the whole candidate set forms one group, as in the single-line schedule of \Cref{sec:example-batch-scheduling}, where the tasks are the ten batches rather than the joined item-batch rows. Their operands are numeric (\kw{INTEGER} or \kw{CONTINUOUS}) or temporal (\kw{TIMESTAMP} with \kw{INTERVAL}) values with $d \ge 0$, taken by position rather than through a dedicated time type. Unlike \querylang{}'s other constructs, these are not SQL features: SQL has no such global constraint, its closest relative being the \kw{OVERLAPS} predicate over a single pair of periods.

\noindent\textbf{NULLs.} A \kw{NULL} in the data, reached while materializing a coefficient, objective term, or bound, is an error that names the offending rows. SQL aggregates skip \kw{NULL} inputs, but here that would silently drop a variable's term from a constraint and exempt it from the limit the constraint imposes; queries instead supply missing values with \kw{COALESCE} or exclude the rows with a filter. The one \kw{NULL} that is not an error is a window function's value at a boundary (\Cref{sec:window-constraints}), which drops that row's constraint by design.

\subsection{Window Constraints}
\label{sec:window-constraints}

A constraint can relate a candidate's decision to the decisions taken before or after it in an order, through SQL window functions over decision columns. \kw{LAG} and \kw{LEAD} refer to the previous and next row of an \kw{ORDER BY}, and \kw{FIRST\_VALUE} and \kw{LAST\_VALUE} to the first and last. The order is given by the window's \kw{ORDER BY}, and any grouping by the constraint's \kwBY{} clause rather than \kw{PARTITION BY}. A production plan over time, with a per-product schedule ordered by period, exercises each:

\begin{lstlisting}[style=deql]
DECIDE plan
FROM production_plan
DECISION COLUMNS (
    produce   CONTINUOUS BETWEEN 0 AND max_rate,
    inventory CONTINUOUS BETWEEN 0 AND warehouse_cap)
SUBJECT TO
  CONSTRAINT balance:
    inventory = COALESCE(LAG(inventory) OVER (ORDER BY period),
                opening_stock) + produce - demand BY product,
  CONSTRAINT ramp:
    LEAD(produce) OVER (ORDER BY period) - produce <= max_ramp BY product,
  CONSTRAINT opening:
    FIRST_VALUE(inventory) OVER (ORDER BY period) >= safety_stock BY product,
  CONSTRAINT closing:
    LAST_VALUE(inventory) OVER (ORDER BY period) >= safety_stock BY product
MINIMIZE SUM(hold_cost * inventory + run_cost * produce);
\end{lstlisting}

Within each product, rows are taken in \texttt{period} order. \texttt{balance} carries inventory forward: \texttt{LAG(inventory)} is the previous period's stock, and in the first period, where there is none, \kw{COALESCE} supplies \texttt{opening\_stock}. \texttt{ramp} bounds the period-to-period increase through \texttt{LEAD(produce)}, which is \kw{NULL} in the last period and drops \texttt{ramp} there. \texttt{opening} and \texttt{closing} hold the safety stock at the first and last periods, through \kw{FIRST\_VALUE} and \kw{LAST\_VALUE}.

A window function carries an \kw{ORDER BY} and no frame (\kw{ROWS} or \kw{RANGE}) clause, and \kw{FIRST\_VALUE} and \kw{LAST\_VALUE} range over the whole ordered group. \querylang{}'s principle is to follow SQL's window semantics; for \kw{LAST\_VALUE} it deliberately departs from SQL. Under SQL's default frame an ordered window ends at the current row, so a frameless \kw{LAST\_VALUE} would return that row rather than the group's last. Because a constraint like \texttt{closing} means the final period's stock, \querylang{} fixes the window to the whole group, making \kw{LAST\_VALUE} the last row of the order; \kw{FIRST\_VALUE} already matches SQL, whose default frame begins at the first row. \kw{LAG} and \kw{LEAD} default to \kw{NULL}, which drops the boundary row's constraint, or to an explicit 0, which retains it. Windowed aggregates such as \texttt{SUM(...) OVER (ORDER BY ...)} are also available for running totals.

\section{Multi-Granularity Decisions}
\label{sec:mg-decisions}

\label{sec:multi-granularity}

Many problems decide at two levels of granularity at once. Batch manufacturing is typical: over a fixed set of items and batches, the plan both assigns items to production batches (a choice per item-batch pair) and schedules when each batch starts (a choice per batch, expressed in minutes from the start of the day). The two decisions share the same batches and are solved together: a batch cannot start before the items placed in it are ready, so which items go into a batch determines how early it can start.

\kwBY{} alone already expresses this, by folding both levels into one candidate set:
\begin{lstlisting}[style=deql]
CREATE CANDIDATES assignments
DECISION KEY (item_id, batch_id) AS
  SELECT i.item_id, b.batch_id, i.weight, i.ready_time,
         b.capacity, b.processing_time
  FROM items i CROSS JOIN batches b;

DECIDE schedule
FROM assignments
DECISION COLUMNS (
    assigned   SELECTION BINARY,
    start_time CONTINUOUS BETWEEN 0 AND 1440 BY batch_id)
SUBJECT TO
  CONSTRAINT one_batch: SUM(assigned) = 1 BY item_id,
  CONSTRAINT batch_cap: SUM(weight * assigned) <= capacity BY batch_id,
  CONSTRAINT ready: start_time >= ready_time * assigned
MINIMIZE MAX(start_time + processing_time);
\end{lstlisting}
This is expressible, but folding both levels into one candidate set repeats each batch once per item. With 100 items and 10 batches, \texttt{assignments} has 1000 rows, and each batch's \texttt{capacity} and \texttt{processing\_time} are repeated on the 100 rows sharing its \texttt{batch\_id}. There are only 10 batch-level start times, yet \texttt{start\_time} must be declared on all 1000 rows and coarsened back to one variable per batch with \kwBY{}. The more columns a batch carries, the more of the query is spent repeating them and collapsing them back. Folding also changes what aggregates compute. Suppose each batch carried a setup cost of 500. On the folded set that cost is a column of a candidate keyed \texttt{(item\_id, batch\_id)}, so \texttt{SUM(setup\_cost)} adds one term per pair (\Cref{sec:variable-grain}): a batch's 100 rows charge it 100 times, 50000 in place of 500. The usual repair is to pre-scale the value, writing 5 in place of 500 so that the 100 row-wise terms add back to 500. The repair holds only while the sum sees all 100 rows. A \kw{WHERE} or a \kwFILTER{} that keeps 60 of them charges 300, and nothing in the query says so. Here not even that holds: \texttt{assigned} is a \kw{SELECTION} column, so auto-multiply (\Cref{sec:package-semantics}) reads the cost as \texttt{SUM(setup\_cost * assigned)}, and the surviving terms are the ones the solver picks, charging a batch that holds 30 items 150 rather than 500.

\querylang{} instead lets each level be its own candidate set, joined on the shared key:
\begin{lstlisting}[style=deql]
CREATE CANDIDATES assignments
DECISION KEY (item_id, batch_id) AS
  SELECT i.item_id, b.batch_id, i.weight, i.ready_time
  FROM items i CROSS JOIN batches b;

CREATE CANDIDATES timing
DECISION KEY (batch_id) AS
  SELECT batch_id, capacity, processing_time FROM batches;

DECIDE schedule
FROM assignments a JOIN timing t ON a.batch_id = t.batch_id
DECISION COLUMNS (
    assigned   SELECTION BINARY ON a,
    start_time CONTINUOUS BETWEEN 0 AND 1440 ON t)
SUBJECT TO
  CONSTRAINT one_batch: SUM(assigned) = 1 BY item_id,
  CONSTRAINT batch_cap: SUM(weight * assigned) <= capacity BY batch_id,
  CONSTRAINT ready: start_time >= ready_time * assigned
MINIMIZE MAX(start_time + processing_time);
\end{lstlisting}
\noindent\textbf{ON clause.} Each decision column in a joined \kwDECIDE{} carries an \kw{ON} tag naming, through its \kw{FROM} alias, the candidate table it belongs to: \texttt{assigned} is decided on \texttt{a}, \texttt{start\_time} on \texttt{t}. The tag sets the column's default grain, the named table's \kwDECISIONKEY{}; an optional \kwBY{}, written before the tag, may coarsen that grain using only columns the same key functionally determines (\Cref{sec:variable-grain}). The tag reuses the \kw{ON} keyword but is not a join condition: a join's \kw{ON} states how rows match, a decision column's \kw{ON} names the table it is decided over. The tag is required on every decision column the query declares whenever \kw{FROM} names more than one candidate table, and an untagged declaration there is rejected (\Cref{sec:validity}); a column inherited from a schema-level \kwDECISIONCOLUMNS{} (\Cref{sec:schema-level}) needs no tag, since it already belongs to the table that declares it. A data column needs no tag either: it carries the grain of the candidate table it comes from, as in the facility-location example (\Cref{sec:example-facility-location}).

\noindent\textbf{Join form and names.} The \kw{JOIN} is an inner join, the only form the grammar admits; an outer join's unmatched rows would carry \kw{NULL}s, which the materialization rules reject (\Cref{sec:constraint-syntax}). Its condition must equate the coarser candidate set's full \kwDECISIONKEY{} with columns of the finer set, as \texttt{a.batch\_id = t.batch\_id} does, so that each row of the finer set matches at most one row of the coarser one; a condition that fans rows out would multiply the terms of every aggregate that reads them, and is rejected (\Cref{sec:validity}). Column names resolve as in SQL: a name found in only one joined table may stay unqualified, and a name found in both must carry a \kw{FROM} alias. The exception is a join-key column equated under the same name on both sides, as \texttt{batch\_id} is: equal on every row, it counts as one column and needs no alias, the treatment SQL gives \kw{USING}. \texttt{batch\_cap} relies on this when it groups \kwBY{}~\texttt{batch\_id}.

The join sets which candidate rows the query ranges over; each decision column's grain sets how many variables it expands into. The join is computed over the materialized candidate relations before the solver runs, so it creates no variables and duplicates none; it can only remove them, since a grain group left with no surviving rows gets no variable, the same rule that applies under \kw{WHERE} (\Cref{sec:variable-grain}). With 100 items and 10 batches the query ranges over 1000 candidate rows and has 1000 \texttt{assigned} variables, one per row of \texttt{assignments}, but only 10 \texttt{start\_time} variables, one per row of \texttt{timing}. A reader coming from SQL might expect 1000 start times, since every joined row shows a start time. Aggregates instead follow the variable count, not the row count: \texttt{SUM(start\_time)} would have 10 terms, one per batch, not 1000. In modeling-language terms, \texttt{start\_time} is indexed by the batches alone, \texttt{var start\_time\{Batches\}} rather than \texttt{var start\_time\{Items, Batches\}}; a \kwDECISIONKEY{} grains one candidate set (\Cref{sec:create-candidates}), and joining candidate sets is how one query carries several such index sets at once. A sparse family needs no new construct: the index set is whatever the defining query returns, so the predicate in a \kwCREATECANDIDATES{} restricts it, as \texttt{worker\_shifts} does (\Cref{sec:example-assignment}).

SQL never needed a tag like \kw{ON}, because the question the tag settles does not arise for data columns. A batch's \texttt{capacity} is a single value that the join copies into every row of the batch, and copying changes nothing: the batch still has one capacity. (How many rows an aggregate adds up is a separate question, settled by grain; \Cref{sec:variable-grain}.) For a decision column the question is real. If each of a batch's 100 rows carried its own start time, the batch could begin at a different moment for each of its items; if the rows share one \texttt{start\_time}, the batch begins once and every item inherits that moment. The result relation cannot settle this, since 100 separate variables may all take equal values. The \kw{ON} tag states which reading is meant.

The \texttt{ready} constraint is where the two levels meet. It is written out rather than generated: interval linking (\Cref{sec:package-semantics}) could produce it only from a declared lower bound of \texttt{ready\_time} on \texttt{start\_time}, and at the batch grain that declaration is invalid, since a coarse variable's bounds must be single-valued within each group (\Cref{sec:grammar-types}) while \texttt{ready\_time} varies across a batch's items. It carries no \kwBY{}, so it instantiates once per candidate row: for a pair \texttt{(i, b)} it reads item \texttt{i}'s \texttt{ready\_time} and batch \texttt{b}'s \texttt{start\_time}. The factor \texttt{assigned} makes it bind only on pairs the plan uses: with \texttt{assigned} $= 0$ it reduces to \texttt{start\_time} $\geq 0$, which the column's lower bound already grants. And because every row of batch \texttt{b} reads the same \texttt{start\_time} variable, the constraint pushes the batch's single start past the ready time of every item placed in it.

The query still produces a single result relation, \texttt{schedule}, not two: the join of the candidate sets with the decision columns filled in, one row per selected item-batch pair (\Cref{sec:package-semantics}). With \texttt{one\_batch} choosing one batch per item, that is 100 rows rather than 1000. In the result each batch's \texttt{start\_time} repeats across the batch's rows exactly as \texttt{capacity} does: the ten variables are printed wherever they apply. A batch that receives no items contributes no rows, so its \texttt{start\_time} does not appear in \texttt{schedule}; reading every batch's start back means joining \texttt{schedule} to \texttt{timing}. Like any \kwDECIDE{} result, \texttt{schedule} is an ordinary relation that further SQL or decision queries can read.

Keeping the levels apart has three advantages over folding them together. A batch's data and decisions live in their own table, one row per batch, so nothing is repeated and no \kwBY{} is needed to recover the batch grain. A batch-level quantity sits at its own grain, so every aggregate that reads it charges it once whenever the batch survives the query's \kw{WHERE}, and a \kwFILTER{} finer than the batch grain, which would keep only some of a batch's rows, is rejected rather than silently rescaling it (\Cref{sec:validity}). And the engine can treat the two separately keyed levels as separate blocks when it selects a formulation (\Cref{sec:pipeline}).

\noindent\textbf{FOREIGN DECISION KEY.} A \kw{FOREIGN DECISION KEY} declares at the schema level the relationship a joined \kwDECIDE{} states per query: the finer set's key columns reference the coarser set's \kwDECISIONKEY{}.
\begin{lstlisting}[style=deql]
CREATE CANDIDATES item_batch_assignments
DECISION KEY (item_id, batch_id)
FOREIGN DECISION KEY (batch_id) REFERENCES batch_timing(batch_id)
AS SELECT ...;
\end{lstlisting}
This is the schema-level counterpart of the join above; it enables the system to validate join structure and identify subproblems it can solve independently. The join itself stays explicit: declaring the foreign key does not let a \kwDECIDE{} omit it.

\noindent\textbf{Design principle.} Decision variables must not appear in \kw{JOIN} conditions; they belong in \kw{SUBJECT TO}. \kw{JOIN} links candidate tables by data columns; constraints link their decision variables. The rule restates the timing above: the join is evaluated before the solver runs, when a decision column has no value to test.

\section{Validity and Errors}
\label{sec:validity}

Validity has a static part, checked on the query alone, and a data-dependent part, checked once the candidate data is read; a key with a duplicate value, or a \kw{NULL} coefficient, cannot be caught earlier. A conforming \querylang{} implementation enforces both, along three lines. \emph{First, it rejects what cannot be given a meaning.} Three checks share one pattern, the single-value rule SQL applies to non-grouped columns under \kw{GROUP BY}: the \kwDECISIONKEY{} must functionally determine every candidate column (\Cref{sec:create-candidates}), a grouped constraint's bound must be single-valued within each group (\Cref{sec:constraint-syntax}), and a coarse variable's bounds must be single-valued within each grain group (\Cref{sec:grammar-types}). When \kw{FROM} names more than one candidate table, every decision column the query declares must carry \kw{ON}, and its \kwBY{} may use only columns the named table's \kwDECISIONKEY{} determines (\Cref{sec:multi-granularity}). The join condition must equate the coarser set's full \kwDECISIONKEY{}, so that no row fans out, and an unqualified column name carried by more than one joined table is rejected unless the join equates it under that name (\Cref{sec:multi-granularity}). A \kw{NULL} reaching a grain column, coefficient, or bound is rejected for the same reason; and every constraint, and the objective when one is given, must reference at least one decision variable once the \kw{SELECTION} desugaring of \Cref{sec:package-semantics} has been applied, since one that does not asserts a fact about the data rather than constraining the decision.

\emph{Second, it rejects, in this version, combinations whose intended meaning is not yet specified.} For instance, the language rejects a \kwFILTER{} predicate that reads a column finer than the aggregated expression's grain, because the union-grain rule would read it as a per-row count rather than as the intended test for any matching row. Precompute a same-grain flag in \kwCREATECANDIDATES{} instead. A later revision may define and admit such cases, and restricting them now keeps existing queries valid when it does.

\emph{Third, patterns that are well-defined but sometimes unintended execute with their documented meaning.} Diagnostics flag the likely mistakes: a row-grain data column multiplying a coarse variable (\Cref{sec:variable-grain}), or a coarse variable that spans several of a grouped constraint's groups and is charged in full against each.

\section{Extensions}
\label{sec:extensions}

\querylang{} provides optional extensions for three needs beyond the core: optimization under uncertainty, inline model scoring, and time- and quality-bounded solving.

\subsection{Stochastic Constraints}
\label{sec:stochastic-constraints}

Following SPaQL~\cite{spaql}, \querylang{} supports stochastic constraints for optimization under uncertainty. The GPU allocation query of \Cref{sec:example-gpu-qos} treats each workload's demand as known; in practice it is uncertain. Uncertain data columns are declared with the \kw{STOCHASTIC} function in \kwCREATECANDIDATES{}, specifying a distribution family and parameters drawn from base table columns:

\begin{lstlisting}[style=deql]
CREATE CANDIDATES gpu_assignments
DECISION KEY (pool_id, workload_id) AS
  SELECT a.pool_id, a.workload_id,
         p.capacity, a.cost, a.latency_ms,
         STOCHASTIC('normal', w.demand_mean, w.demand_std)
           AS demand
  FROM assignment_options a
  JOIN gpu_pools p ON a.pool_id = p.pool_id
  JOIN workloads w ON a.workload_id = w.workload_id;
\end{lstlisting}

Constraints over stochastic columns use \kw{WITH PROBABILITY} to specify chance constraints, and the objective can use \kw{EXPECTED} to optimize in expectation:

\begin{lstlisting}[style=deql]
DECIDE allocation_plan
FROM gpu_assignments
DECISION COLUMNS (
    active   SELECTION BINARY,
    quantity CONTINUOUS BETWEEN 0 AND capacity)
WHERE latency_ms <= 200
SUBJECT TO
  CONSTRAINT pool_capacity:
    SUM(quantity) <= capacity BY pool_id,
  CONSTRAINT meet_demand:
    SUM(quantity) >= demand BY workload_id
    WITH PROBABILITY >= 0.95
MINIMIZE EXPECTED SUM(cost * quantity);
\end{lstlisting}

The \kw{WITH PROBABILITY} clause on a constraint enforces it as a chance constraint: the system samples scenarios from the declared distributions and ensures the constraint holds with at least the specified probability. \kw{EXPECTED} on the objective optimizes the expected value across scenarios.

\subsection{Inline Model Scoring}
\label{sec:predict}

\kwPREDICT{} invokes a registered model during candidate construction, inside the \kw{SELECT} of \kwCREATECANDIDATES{}, to populate a data column from other columns. The model may be classical ML or an LLM; \kw{PREDICT}(\textit{model}, \textit{args}) returns a scalar that then participates in constraints and objectives like any other data column, following the \kw{PREDICT} pattern of BigQuery ML~\cite{bigquery-ml}. The budget-allocation example (\Cref{sec:example-budget-allocation}) scores expected ROI this way.

\subsection{Quality Modifiers}
\label{sec:quality-modifiers}

Quality modifiers \kw{WITHIN} and \kw{TIMEOUT} on the objective clause grant the decision planner permission to trade solution quality for speed. Two modifiers are supported:

\begin{lstlisting}[style=deql]
MINIMIZE SUM(cost * quantity) WITHIN 5% TIMEOUT 30s
\end{lstlisting}

\kw{WITHIN} $X$\kw{\%} asks the planner to stop once the solution is provably within $X$\% of optimal; on its own, it bounds the optimality gap. \kw{TIMEOUT} $X$\kw{s} caps solve time, returning the best feasible solution found within $X$ seconds. Both modifiers are optional and compose: given both, the planner stops at whichever bound is reached first and reports the achieved gap, so a timeout that fires before the gap is proven returns the best solution found rather than a gap-bounded one. They enable the decision planner to select abstraction-based decomposition or approximation algorithms when solving exactly is impractical for large instances (\Cref{sec:pipeline}).

\section{Query Examples}
\label{sec:query-examples}

\subsection{Subset Selection (Knapsack)}
\label{sec:example-knapsack}

Select products for a shopping cart, maximizing total rating subject to budget and weight limits:
\begin{lstlisting}[style=deql]
CREATE CANDIDATES products
DECISION KEY (product_id) AS
  SELECT product_id, price, weight, rating, category
  FROM StockItems WHERE in_stock = TRUE;

DECIDE cart
FROM products
DECISION COLUMNS (chosen SELECTION BINARY)
SUBJECT TO
  CONSTRAINT budget:       SUM(price) <= 100,
  CONSTRAINT total_weight: SUM(weight) <= 50,
  CONSTRAINT per_category: SUM(weight) <= 20 BY category
MAXIMIZE SUM(rating);
\end{lstlisting}

\subsection{Continuous Allocation (Diet Problem)}
\label{sec:example-diet}

Determine servings of each food to meet nutritional requirements at minimum cost. Each nutrient is a column of the candidate, so \texttt{(item\_id)} is a key of the data and the cost objective counts each food once:
\begin{lstlisting}[style=deql]
CREATE CANDIDATES diet_options
DECISION KEY (item_id) AS
  SELECT item_id, unit_cost, protein, fiber
  FROM Foods;

DECIDE diet_plan
FROM diet_options
DECISION COLUMNS (servings CONTINUOUS BETWEEN 0 AND 10)
SUBJECT TO
  CONSTRAINT min_protein: SUM(protein * servings) >= 50,
  CONSTRAINT max_protein: SUM(protein * servings) <= 160,
  CONSTRAINT min_fiber:   SUM(fiber * servings) >= 25
MINIMIZE SUM(unit_cost * servings);
\end{lstlisting}
A normalized variant keeps nutrients as rows of a separate table at the \texttt{(item\_id, nutrient)} grain; coupling that finer table with per-item servings joins two grains in one query (\Cref{sec:multi-granularity}).

\subsection{Mixed Integer (Production Planning)}
\label{sec:example-production}

Choose production quantities for each product to meet demand at minimum cost, with integer production levels:
\begin{lstlisting}[style=deql]
CREATE CANDIDATES production_options
DECISION KEY (product_id) AS
  SELECT product_id, production_cost, demand,
         min_production, max_production
  FROM Products;

DECIDE production_plan
FROM production_options
DECISION COLUMNS (
    units_produced INTEGER
      BETWEEN min_production AND max_production)
SUBJECT TO
  CONSTRAINT meet_demand:
    units_produced >= demand
MINIMIZE SUM(production_cost * units_produced);
\end{lstlisting}

\subsection{Assignment (Worker-Shift Matching)}
\label{sec:example-assignment}

Assign workers to shifts such that each worker gets at most one shift and each shift has exactly one worker, minimizing skill mismatch:
\begin{lstlisting}[style=deql]
CREATE CANDIDATES worker_shifts
DECISION KEY (worker_id, shift_id)
DECISION COLUMNS (assigned SELECTION BINARY) AS
  SELECT w.worker_id, s.shift_id,
         ABS(w.skill_level - s.required_skill) AS mismatch
  FROM Workers w CROSS JOIN Shifts s
  WHERE w.skill_level >= s.required_skill;

DECIDE schedule
FROM worker_shifts
SUBJECT TO
  CONSTRAINT one_shift:  COUNT(*) <= 1 BY worker_id,
  CONSTRAINT one_worker: COUNT(*) = 1 BY shift_id
MINIMIZE SUM(mismatch);
\end{lstlisting}

The composite \kwDECISIONKEY{} \texttt{(worker\_id, shift\_id)} with mutual-exclusivity constraints reveals totally unimodular structure (\Cref{sec:pipeline}): the LP relaxation is integral, so the system can solve it with a network-flow method such as network simplex rather than general branch-and-bound.

\subsection{Batch Scheduling (Multi-Granularity)}
\label{sec:example-batch-scheduling}

Assign items to batches while determining each batch's start time: the query of \Cref{sec:multi-granularity} extended with a per-item deadline and a no-overlap constraint, under the table names of \Cref{sec:schema-level}. Two candidate tables carry different \kwDECISIONKEY{}s:
\begin{lstlisting}[style=deql]
CREATE CANDIDATES item_batch_assignments
DECISION KEY (item_id, batch_id) AS
  SELECT i.item_id, b.batch_id,
         i.weight, i.deadline
  FROM items i CROSS JOIN batches b;

CREATE CANDIDATES batch_timing
DECISION KEY (batch_id) AS
  SELECT batch_id, capacity, processing_time
  FROM batches;

DECIDE schedule
FROM item_batch_assignments a
  JOIN batch_timing t ON a.batch_id = t.batch_id
DECISION COLUMNS (
    assigned   SELECTION BINARY ON a,
    start_time CONTINUOUS BETWEEN 0 AND 1440 ON t)
SUBJECT TO
  CONSTRAINT one_batch:
    SUM(assigned) = 1 BY item_id,
  CONSTRAINT batch_cap:
    SUM(weight * assigned) <= t.capacity BY batch_id,
  CONSTRAINT item_deadline:
    start_time + processing_time
      <= deadline + (1440 + processing_time - deadline) * (1 - assigned),
  CONSTRAINT sequence:
    DISJUNCTIVE(start_time, processing_time)
MINIMIZE MAX(start_time + processing_time);
\end{lstlisting}
The \texttt{item\_deadline} constraint binds only the pair the plan uses. With \texttt{assigned} $= 0$ the added term equals the pair's own slack against \texttt{start\_time}'s upper bound, so the constraint reduces to \texttt{start\_time} $\leq 1440$ and says nothing; a round constant such as 1440 would be too small whenever an item's \texttt{deadline} falls below a batch's \texttt{processing\_time}, and would then cap the start of a batch that does not hold the item. The gated lower bound \texttt{ready} of \Cref{sec:multi-granularity} needs no such term, since with \texttt{assigned} $= 0$ it relaxes to the variable's own lower bound. \texttt{sequence} carries no \kwBY{}, so all batches form one group: a single production line.

\subsection{Budget Allocation (ML Scoring)}
\label{sec:example-budget-allocation}

Allocate marketing spend across products at discrete levels, where expected ROI is predicted by an ML model:
\begin{lstlisting}[style=deql]
CREATE CANDIDATES spend_options
DECISION KEY (product_id, spend_level)
DECISION COLUMNS (
  chosen SELECTION BINARY)
CONSTRAINTS (
  one_level: COUNT(*) = 1 BY product_id)
AS SELECT p.product_id, l.spend_level, l.spend,
         PREDICT(roi_model, p.product_id, l.spend_level)
           AS expected_roi
  FROM Products p
  CROSS JOIN SpendLevels l;

DECIDE spending_plan
FROM spend_options
SUBJECT TO
  CONSTRAINT total_budget: SUM(spend) <= 1000000
MAXIMIZE SUM(expected_roi);
\end{lstlisting}

The \kw{PREDICT} function invokes a registered ML model inline. The \kw{SELECTION} modifier on \texttt{chosen} ensures auto-multiply applies to \texttt{SUM(spend)} and \texttt{SUM(expected\_roi)}. The schema-level constraint \texttt{one\_level} ensures exactly one spend level is selected per product.

\subsection{Facility Location (Multi-Granularity)}
\label{sec:example-facility-location}

Decide which facilities to open (facility-level binary decision) and how much demand each facility serves for each customer (facility-customer-level continuous decision):
\begin{lstlisting}[style=deql]
CREATE CANDIDATES facility_options
DECISION KEY (facility_id) AS
  SELECT facility_id, open_cost, capacity
  FROM Facilities;

CREATE CANDIDATES facility_assignments
DECISION KEY (facility_id, customer_id) AS
  SELECT f.facility_id, c.customer_id,
         c.demand, t.transport_cost
  FROM Facilities f
  CROSS JOIN Customers c
  JOIN TransportCosts t
    ON t.facility_id = f.facility_id
    AND t.customer_id = c.customer_id;

DECIDE location_plan
FROM facility_options f
  JOIN facility_assignments a
    ON f.facility_id = a.facility_id
DECISION COLUMNS (
    opened BINARY ON f,
    amount CONTINUOUS BETWEEN 0 AND demand ON a)
SUBJECT TO
  CONSTRAINT serve:
    SUM(amount) = demand BY customer_id,
  CONSTRAINT facility_cap:
    SUM(amount) <= capacity * opened BY facility_id
MINIMIZE SUM(open_cost * opened)
       + SUM(transport_cost * amount);
\end{lstlisting}

The \texttt{opened} variable is indexed by \texttt{facility\_id}; the \texttt{amount} variable is indexed by \texttt{(facility\_id, customer\_id)}. The constraint \texttt{SUM(amount) <= capacity * opened BY facility\_id} links the two levels: a closed facility serves no demand. The objective term \texttt{SUM(open\_cost * opened)} is grain-scoped (\Cref{sec:variable-grain}): each facility's fixed cost is counted once, not once per customer it serves, because \texttt{open\_cost} lives on the facility-grain \texttt{facilities} candidate. Carried instead on the finer \texttt{facility\_assignments}, the same \texttt{SUM} would bill it once per customer served.

\subsection{Multi-Resource Allocation with QoS (GPU Scheduling)}
\label{sec:example-gpu-qos}

Allocate GPU-hours from pools to workloads, respecting two resources (compute and memory) and a quality-of-service reservation. The candidate set pairs each workload with the pools that can host it, and the three filtering levels each play a distinct role.
\begin{lstlisting}[style=deql]
CREATE CANDIDATES gpu_assignments
DECISION KEY (pool_id, workload_id) AS
  SELECT a.pool_id, a.workload_id,
         p.capacity, p.memory_gb,
         w.demand, w.qos_class, w.mem_per_unit,
         a.cost, a.latency_ms
  FROM assignment_options a
  JOIN gpu_pools p ON a.pool_id = p.pool_id
  JOIN workloads w ON a.workload_id = w.workload_id
  WHERE p.memory_gb >= w.min_memory_gb;

DECIDE allocation_plan
FROM gpu_assignments
DECISION COLUMNS (
    active   SELECTION BINARY,
    quantity CONTINUOUS BETWEEN 0 AND capacity)
WHERE latency_ms <= 200
SUBJECT TO
  CONSTRAINT pool_capacity:
    SUM(quantity) <= capacity BY pool_id,
  CONSTRAINT mem_capacity:
    SUM(mem_per_unit * quantity) <= memory_gb BY pool_id,
  CONSTRAINT meet_demand:
    SUM(quantity) = demand BY workload_id,
  CONSTRAINT sla_reserve:
    SUM(quantity) FILTER (WHERE qos_class = 'realtime')
      >= 0.4 * capacity BY pool_id
MINIMIZE SUM(cost * quantity);
\end{lstlisting}

Three filtering levels appear, each with a distinct role. \kw{WHERE} in \kwCREATECANDIDATES{} (\texttt{memory\_gb >= min\_memory\_gb}) removes pairings that are infeasible regardless of the decision (a pool with too little memory can never host the workload), so they never enter the candidate set. \kw{WHERE} in \kwDECIDE{} (\texttt{latency\_ms <= 200}) applies a per-query policy filter that can change between runs without redefining the candidates (\Cref{sec:decide-statement}). \kwFILTER{} inside \texttt{sla\_reserve} scopes a single aggregate to realtime workloads without removing any candidate from the problem (\Cref{sec:constraint-syntax}): an excluded candidate still participates in \texttt{pool\_capacity}, \texttt{mem\_capacity}, and \texttt{meet\_demand}. The two resources, compute (\texttt{pool\_capacity}) and memory (\texttt{mem\_capacity}), are independent constraints grouped \kwBY{} \texttt{pool\_id}; a further resource is a further constraint, not a new construct.

\section{Formal Grammar}
\label{sec:grammar}

This section presents the formal BNF grammar for \querylang{}. Terminal symbols appear in \texttt{UPPERCASE} and non-terminals as lowercase names in angle brackets (\texttt{<name>}); square brackets mark optional elements, \texttt{|} separates alternatives, and \texttt{\{ \}*} marks repetition. The leaf non-terminals \texttt{<identifier>}, \texttt{<alias>}, \texttt{<literal>}, \texttt{<number>}, \texttt{<string\_literal>}, \texttt{<predicate>}, and \texttt{<select\_query>} carry their standard SQL meanings, and \texttt{<expr>} and \texttt{<scalar\_expr>} are used interchangeably for a scalar SQL expression over data columns, decision columns, and literals, arithmetic included (e.g., \texttt{start\_time + processing\_time}). The function forms \texttt{<window\_fn>}, \texttt{<stochastic\_expr>}, and \texttt{<predict\_expr>} defined below extend \texttt{<expr>}, each usable wherever a scalar expression is allowed in the context noted with it. Parentheses group alternatives within a production. Parentheses that belong to the syntax itself, as in \kw{SUM} \texttt{( <expr> )}, are terminals. The time-unit suffixes \texttt{s}, \texttt{ms}, and \texttt{m} are terminals too, as is the percent sign. Every statement ends with a semicolon.

\subsection{CREATE CANDIDATES}
\label{sec:grammar-candidates}

\begin{small}
\begin{verbatim}
<create_candidates_stmt> ::=
    CREATE CANDIDATES <identifier>
    DECISION KEY ( <column_list> )
    [ FOREIGN DECISION KEY ( <column_list> )
      REFERENCES <identifier> ( <column_list> )
      [ AS <alias> ] ]
    [ DECISION COLUMNS ( <decision_decl_list> ) ]
    [ CONSTRAINTS ( <schema_constraint_list> ) ]
    AS <select_query>

<column_list> ::=
    <identifier> { , <identifier> }*

<decision_decl_list> ::=
    <decision_decl> { , <decision_decl> }*

<decision_decl> ::=
    <identifier> [ SELECTION ] <var_type> [ <bounds> ]
    [ <by_clause> ]
    [ ON <alias> ]
  | <identifier> = <aggregate_expr>
    [ REFERENCES <identifier> ( <column_list> ) ]

<var_type> ::=
    BINARY | INTEGER | CONTINUOUS

<bounds> ::=
    BETWEEN <bound> AND <bound>

<bound> ::=
    <expr> | UNBOUNDED

<by_clause> ::=
    BY <column_ref>
  | BY ( <column_ref> { , <column_ref> }* )
  | BY ( )

<column_ref> ::=
    [ <alias> . ] <identifier>
\end{verbatim}
\end{small}

\subsection{DECIDE Statement}
\label{sec:grammar-decide}

\begin{small}
\begin{verbatim}
<decide_stmt> ::=
    DECIDE [ [ INTO ] <identifier> ]
    FROM <candidate_ref> { <join_clause> }*
    [ DECISION COLUMNS ( <decision_decl_list> ) ]
    [ WHERE <predicate> ]
    SUBJECT TO <constraint_list>
    [ <objective_clause> ]

<candidate_ref> ::=
    <identifier> [ <alias> ]
  | <table_or_query> DECISION KEY ( <column_list> ) [ <alias> ]

<table_or_query> ::=
    <identifier> | ( <select_query> )

<join_clause> ::=
    JOIN <candidate_ref> ON <join_condition>

<join_condition> ::=
    <column_ref> = <column_ref>
      { AND <column_ref> = <column_ref> }*

<objective_clause> ::=
    ( MINIMIZE | MAXIMIZE ) [ EXPECTED ] <objective_expr>
    [ WITHIN <number> % ]
    [ TIMEOUT <number> ( s | ms | m ) ]

<objective_expr> ::=
    <objective_term> { ( + | - ) <objective_term> }*

<objective_term> ::=
    [ <number> * ] <aggregate_expr>
  | <scalar_expr>

<with_stmt> ::=
    WITH <with_step> { , <with_step> }*
    <select_query>

<with_step> ::=
    <identifier> AS ( <select_query> )
  | <identifier> AS ( <decide_stmt> )
\end{verbatim}
\end{small}

A \kwDECIDE{} statement may also appear as a named step in a SQL \kw{WITH} clause, which \texttt{<with\_stmt>} covers: used this way the \kwDECIDE{} omits the \kw{INTO} \texttt{<identifier>} result name and produces a relation under the step's name, composing with SQL CTEs and queries (\Cref{sec:composability}). A step that is an ordinary \texttt{<select\_query>} keeps its standard SQL meaning; only a step holding a \kwDECIDE{} is new. The \kw{ON} tag on a decision column is optional in the grammar but required whenever \kw{FROM} names more than one candidate table (\Cref{sec:multi-granularity}). A \texttt{<column\_ref>} resolves as in SQL: its alias qualifier is required when the name is carried by more than one joined candidate table and the join does not equate it under that name (\Cref{sec:multi-granularity}). A \texttt{<join\_condition>} is a conjunction of column equalities; \Cref{sec:multi-granularity} further requires it to equate the coarser candidate set's full \kwDECISIONKEY{} with columns of the finer set.

\subsection{Constraints}
\label{sec:grammar-constraints}

\begin{small}
\begin{verbatim}
<constraint_list> ::=
    <constraint_def> { , <constraint_def> }*

<constraint_def> ::=
    [ CONSTRAINT <identifier> : ] <constraint_body>

<constraint_body> ::=
    <aggregate_constraint>
  | <per_row_constraint>
  | <scheduling_constraint>

<aggregate_constraint> ::=
    <aggregate_expr> [ FILTER ( WHERE <predicate> ) ]
    <comparison_op> <bound_expr>
    [ <by_clause> ]
    [ WITH PROBABILITY <comparison_op> <number> ]

<per_row_constraint> ::=
    <scalar_expr> <comparison_op> <scalar_expr>
    [ <by_clause> ]

<scheduling_constraint> ::=
    DISJUNCTIVE ( <column_ref> , <column_ref> )
      [ <by_clause> ]
  | CUMULATIVE ( <column_ref> , <column_ref> ,
      <column_ref> ) <comparison_op> <expr>
      [ <by_clause> ]

<comparison_op> ::=
    <= | >= | = | < | >

<aggregate_expr> ::=
    SUM ( <expr> )
  | COUNT ( * )
  | MIN ( <expr> )
  | MAX ( <expr> )
  | AVG ( <expr> )

<bound_expr> ::=
    <literal> | <identifier> | <scalar_expr>

<window_fn> ::=
    ( LAG | LEAD ) ( <expr> [ , <number> [ , <expr> ] ] )
      OVER ( ORDER BY <column_list> )
  | ( FIRST_VALUE | LAST_VALUE ) ( <expr> )
      OVER ( ORDER BY <column_list> )
  | <aggregate_expr> OVER ( ORDER BY <column_list> )
\end{verbatim}
\end{small}

The \kwBY{} clause in aggregate constraints generates one constraint instance per distinct value of the grouping columns, with the bound expression evaluated per group; every data column in the bound must be single-valued within its group (the grouping columns must functionally determine it), or the constraint is rejected. The \kwFILTER{} clause restricts which candidate rows participate in the aggregate. In \kw{LAG} and \kw{LEAD} the second argument is the row offset and the third the value substituted at a boundary, as in SQL. Window functions (\kw{LAG}, \kw{LEAD}, \kw{FIRST\_VALUE}, \kw{LAST\_VALUE}) may appear within a constraint expression to relate a decision to others in an \kw{ORDER BY} (\Cref{sec:window-constraints}); they take no \kw{PARTITION BY}, grouping instead by the constraint's \kwBY{} clause.

\subsection{Schema-Level Constructs}
\label{sec:grammar-schema}

\begin{small}
\begin{verbatim}
<schema_constraint_list> ::=
    <schema_constraint> { , <schema_constraint> }*

<schema_constraint> ::=
    <identifier> : <constraint_body>

<create_abstraction_stmt> ::=
    CREATE ABSTRACTION <identifier>
    ON <identifier> ( <column_list> )
    USING <identifier> ( [ <param_list> ] )

<param_list> ::=
    <param> { , <param> }*

<param> ::=
    <identifier> = <expr>
\end{verbatim}
\end{small}

Schema-level \kwDECISIONCOLUMNS{} and \kw{CONSTRAINTS} are declared inside \kwCREATECANDIDATES{} and automatically enforced in any \kwDECIDE{} query referencing those candidates. Decision columns may include derived columns using \texttt{=} syntax with an optional \kw{REFERENCES} clause for cross-level aggregation in multi-granularity problems. The \kwCREATEABSTRACTION{} statement defines a coarser, cluster-based view of a candidate set for decomposition-based solving (\Cref{sec:abstraction,sec:pipeline}); the \kw{USING} clause names a clustering method and its parameters.

\subsection{Stochastic Constructs}
\label{sec:grammar-stochastic}

\begin{small}
\begin{verbatim}
<stochastic_expr> ::=
    STOCHASTIC ( <string_literal> ,
      <expr> , <expr> )
\end{verbatim}
\end{small}

The \kw{STOCHASTIC} function declares an uncertain data column. The first argument is a distribution family (\texttt{'normal'}, \texttt{'uniform'}); the remaining arguments are distribution parameters drawn from base table columns. \kw{WITH PROBABILITY} on a constraint enforces it as a chance constraint: the system samples scenarios and ensures the constraint holds with at least the specified probability. \kw{EXPECTED} on the objective optimizes the expected value across sampled scenarios.

\subsection{Prediction and AI Column Population}
\label{sec:grammar-predict}

\begin{small}
\begin{verbatim}
<predict_expr> ::=
    PREDICT ( <model_name> , <expr_list> )

<model_name> ::=
    <identifier>

<expr_list> ::=
    <expr> { , <expr> }*
\end{verbatim}
\end{small}

The \kw{PREDICT} function invokes a registered ML or AI model within a \kw{SELECT} clause of \kwCREATECANDIDATES{}. The first argument names a model registered in the system catalog; subsequent arguments are column expressions passed as model inputs. The result is a scalar value that populates a data column participating in constraints and objectives. This covers both traditional ML model scoring (e.g., regression, classification) and LLM-based column population (e.g., sentiment analysis, entity extraction, compatibility scoring).

\subsection{Type System}
\label{sec:grammar-types}

Decision variables have three base types: \kw{BINARY} ($\{0,1\}$), \kw{INTEGER} ($\mathbb{Z}$), and \kw{CONTINUOUS} ($\mathbb{R}$). The \kw{SELECTION} modifier is orthogonal to the base type and adds auto-multiply and interval-linking semantics; it also shapes the result, since a candidate whose selection variable is $0$ is omitted (\Cref{sec:package-semantics}). When \kw{BETWEEN} is omitted, \kw{BINARY} variables range over $\{0,1\}$, and \kw{INTEGER} and \kw{CONTINUOUS} variables default to $[0, +\infty)$.

\begin{table}[t]
\centering
\footnotesize
\renewcommand{\arraystretch}{1.15}
\caption{Type combinations and their semantics. $p$ denotes the selection variable; $x$ denotes a companion variable.}
\label{tab:type-combinations}
\begin{tabular}{@{}llp{6cm}@{}}
\toprule
\textbf{Declaration} & \textbf{Domain} & \textbf{Semantics} \\
\midrule
\kw{BINARY} & $\{0, 1\}$ & Standalone binary decision \\
\kw{INTEGER BETWEEN} $\ell$ \kw{AND} $u$ & $[\ell, u] \cap \mathbb{Z}$ & Bounded integer \\
\kw{CONTINUOUS BETWEEN} $\ell$ \kw{AND} $u$ & $[\ell, u]$ & Bounded continuous \\
\kw{SELECTION BINARY} & $\{0, 1\}$ & Selection indicator with auto-multiply \\
\kw{SELECTION INTEGER BETWEEN} $0$ \kw{AND} $k$ & $\{0, \ldots, k\}$ & Repetition count with auto-multiply \\
$x$ \kw{CONTINUOUS BETWEEN} $\ell$ \kw{AND} $u$ with $p$ \kw{SELECTION BINARY} & $\{0\} \cup [\ell, u]$ & $x$ interval-linked to $p$ ($\ell p \le x \le u p$) at $p$'s grain on $p$'s table \\
\bottomrule
\end{tabular}
\end{table}

The type checker enforces the following rules:
\begin{itemize}
\item Objective and constraint expressions may be linear, quadratic, or nonlinear in decision variables. The engine selects a solver backend based on expression structure (e.g., a general LP or MILP solver for a linear formulation; \Cref{sec:pipeline}); the available backends, and the routing of quadratic and nonlinear formulations, lie outside this specification.
\item \kw{SELECTION} variables must be \kw{BINARY} or \kw{INTEGER} (not \kw{CONTINUOUS}).
\item A candidate table carries at most one \kw{SELECTION} variable.
\item \kw{SELECTION} variables admit no \kwBY{} clause: the desugaring rules of \Cref{sec:package-semantics} are defined per candidate row, so a selection variable stays at the full \kwDECISIONKEY{} grain of its candidate table.
\item \kwBY{} columns must reference data columns (not decision variables).
\item A strict comparison (\texttt{<} or \texttt{>}) is admitted only when every decision variable in the constraint is \kw{BINARY} or \kw{INTEGER}, in which case the engine tightens it to the next integer; over a \kw{CONTINUOUS} variable it leaves an open region with no attainable optimum and is rejected.
\item Either end of \kw{BETWEEN} may be \kw{UNBOUNDED} (the keyword SQL uses in window frames), giving one-sided bounds.
\item Bounds in \kw{BETWEEN} can reference data columns. They are evaluated once per variable instance: per candidate row for full-key variables, and per group for variables with a \kwBY{} grain, in which case every referenced data column must be single-valued within each group (the grain columns must functionally determine it); a declaration whose bound varies within a group is rejected.
\item Interval linking (\Cref{sec:package-semantics}) applies to companions of a \kw{SELECTION BINARY} variable; a \kw{SELECTION INTEGER} variable multiplies aggregates but does not link companions.
\item A linked companion's bounds must satisfy $\ell \le u$; a constant empty band is rejected statically, and a data-valued bound that yields $\ell > u$ (or \kw{NULL}) on some candidate row is rejected at materialization, naming the offending rows (\Cref{sec:validity}).
\item A linked companion whose bound is infinite on either end (defaulted or \kw{UNBOUNDED}) cannot be linked through the generated product; the engine uses an indicator constraint or a finite bound implied by the constraints, and rejects the query only when neither is available. Because bounds may reference data columns, this check is per instance, like the \kw{NULL} rule.
\end{itemize}

\subsection{Reserved Keywords}
\label{sec:grammar-keywords}

The list below covers the keywords \querylang{} introduces and the SQL keywords to which it gives a special meaning; a few are already reserved in standard SQL (e.g., \kw{BETWEEN}, \kw{CONSTRAINT}, \kw{REFERENCES}) and are included for completeness. Standard SQL clause and function keywords that \querylang{} uses unchanged keep their usual meaning and are not repeated below: \kw{SELECT}, \kw{FROM}, \kw{WHERE}, \kw{AS}, \kw{AND}, \kw{CREATE}, \kw{ORDER BY}. The aggregate functions \kw{SUM}, \kw{COUNT}, \kw{MIN}, \kw{MAX}, \kw{AVG} are likewise not listed here, but their grain-based semantics differ from SQL's (\Cref{sec:variable-grain}).

\begin{itemize}
\item \textbf{Statements:} \kw{CANDIDATES}, \kw{DECIDE}, \kw{INTO}
\item \textbf{Decision constructs:} \kw{DECISION}, \kw{KEY}, \kw{COLUMNS}, \kw{CONSTRAINTS}, \kw{SELECTION}
\item \textbf{Variable types:} \kw{BINARY}, \kw{INTEGER}, \kw{CONTINUOUS}
\item \textbf{Variable modifiers:} \kw{BETWEEN}, \kw{UNBOUNDED}
\item \textbf{Objective:} \kw{MINIMIZE}, \kw{MAXIMIZE}
\item \textbf{Constraints:} \kw{SUBJECT TO}, \kw{CONSTRAINT}, \kwBY{}, \kwFILTER{}, \kw{WITH PROBABILITY}
\item \textbf{Scheduling:} \kw{DISJUNCTIVE}, \kw{CUMULATIVE}
\item \textbf{Window functions:} \kw{LAG}, \kw{LEAD}, \kw{FIRST\_VALUE}, \kw{LAST\_VALUE}, \kw{OVER}
\item \textbf{Multi-granularity:} \kw{FOREIGN DECISION KEY}, \kw{REFERENCES}, \kw{JOIN}, \kw{ON}
\item \textbf{Quality modifiers:} \kw{WITHIN}, \kw{TIMEOUT}
\item \textbf{Stochastic (extension):} \kw{STOCHASTIC}, \kw{PROBABILITY}, \kw{EXPECTED}
\item \textbf{Prediction (extension):} \kw{PREDICT}
\item \textbf{Candidate abstractions:} \kw{ABSTRACTION}, \kw{USING}
\end{itemize}

\section{Execution Model}
\label{sec:pipeline}

A \querylang{} query flows through three intermediate representations before producing a result relation.

The \textbf{parser} (a SQL parser extension) tokenizes the query and produces \textbf{Abstract IR}, a normalized, symbolic, database-agnostic representation that preserves column references, symbolic expressions, constraint structure with grouping and filtering, and decision column declarations with their grains, together with each candidate's \kwDECISIONKEY{}. Abstract IR encodes no algorithm choice and binds no data. Expression grains (\Cref{sec:variable-grain}) are inferred on these surface expressions, before any algebraic simplification: rewriting can change a grain. With \texttt{cap} declared \kw{CONTINUOUS} \kwBY{} \texttt{site}, \texttt{SUM(cap)} is one term per site, but \texttt{SUM(cap * flag)} with \texttt{flag} a data column ranges over the rows even when \texttt{flag} always equals 1, so eliminating the factor would change the result.

The \textbf{decision planner} analyzes the Abstract IR through structure detection (e.g., recognizing network flow or total unimodularity from variable-constraint patterns) and formulation selection. The decision planner produces \textbf{Concrete IR}, which encodes the chosen mathematical formulation (still symbolic, using column references rather than actual values). Different formulations produce structurally different Concrete IRs from the same Abstract IR (e.g., generic MILP vs.\ LP relaxation vs.\ min-cost network flow).

The \textbf{materializer} fetches candidate data from the underlying sources, computes the \kw{JOIN}s a \kwDECIDE{} states between candidate tables, and evaluates symbolic expressions against actual rows, producing \textbf{Materialized IR}: an objective coefficient vector, a sparse constraint matrix, variable bounds, and type annotations. This is what the chosen backend receives. The \textbf{executor} dispatches the Materialized IR to a solver or algorithm suited to the formulation (a general-purpose LP/MILP solver for a coefficient matrix, a network-flow algorithm for a flow graph), then joins the resulting variable assignments back with candidate metadata to form the output relation. The language commits to no solver; the planner selects a backend per formulation.

The output is a standard relation that can be queried with SQL, joined with other tables, or fed as input to another \kwDECIDE{} query.

\section{Scope and Limitations}
\label{sec:discussion}

\subsection{Current Scope}
\label{sec:expressiveness}

\querylang{}'s scope follows from its architecture (\Cref{sec:pipeline}). Any problem stated as typed decision variables, constraints, and an objective over candidate relations is captured symbolically in the Abstract IR; finding a formulation and a solver or algorithm for it is the planner's responsibility. The language therefore expresses a range of problems, including:

\begin{itemize}
\item \textbf{Linear decision problems}: subset selection, resource allocation, assignment, and scheduling, expressed with \kw{BINARY}, \kw{INTEGER}, and \kw{CONTINUOUS} decision variables, linear objectives and constraints, and \kw{SELECTION} desugaring.
\item \textbf{Nonlinear and quadratic objectives and constraints}: expressions in which decision variables are multiplied or divided, for example portfolio variance from a covariance term or cannibalization penalties in budget allocation. Expressions may be linear, quadratic, or nonlinear in the decision variables; choosing a solver for them lies outside this specification.
\item \textbf{Constraint programming constructs}: \kw{DISJUNCTIVE} and \kw{CUMULATIVE} for scheduling problems, which desugar to mixed-integer constraints (\Cref{sec:constraint-syntax}).
\item \textbf{Extensions beyond the core}: optimization under uncertainty (\kw{STOCHASTIC}, chance constraints, expected objectives), inline model scoring (\kw{PREDICT}), and time- and quality-bounded solving (\kw{WITHIN}, \kw{TIMEOUT}), each specified in \Cref{sec:extensions}.
\end{itemize}

\subsection{Relationship to PaQL}
\label{sec:paql-compat}

\querylang{} is backward-compatible with PaQL~\cite{paql}: every package query has a \kwDECIDE{} equivalent. PaQL states the options and the package constraints in one statement. The query below builds a meal plan from gluten-free recipes, repeating each recipe up to a limit its own \texttt{max\_servings} sets, to meet a calorie floor at minimum fat; its \querylang{} translation follows.

\noindent The PaQL query is:
\begin{lstlisting}[style=deql]
SELECT PACKAGE(*) AS meal_plan
FROM Recipes
REPEAT max_servings
WHERE gluten = 0
SUCH THAT SUM(kcal) >= 2000
MINIMIZE SUM(fat)
\end{lstlisting}

\noindent The \querylang{} translation is:
\begin{lstlisting}[style=deql]
CREATE CANDIDATES gluten_free_recipes
DECISION KEY (recipe_id) AS
  SELECT recipe_id, kcal, fat, max_servings
  FROM Recipes WHERE gluten = 0;

DECIDE meal_plan
FROM gluten_free_recipes
DECISION COLUMNS (servings SELECTION INTEGER
                    BETWEEN 0 AND max_servings + 1)
SUBJECT TO
  CONSTRAINT calories: SUM(kcal) >= 2000
MINIMIZE SUM(fat);
\end{lstlisting}

Each PaQL clause has a \querylang{} counterpart. \kw{PACKAGE(*)} corresponds to a \kwDECIDE{} over a candidate set; PaQL's tuple-level \kw{WHERE} to the candidate-defining \kw{WHERE}; and each package-level \kw{SUCH THAT} constraint to a named \kw{SUBJECT TO} constraint. \kw{REPEAT} $\rho$, which lets a tuple appear up to $\rho + 1$ times, becomes a \kw{SELECTION INTEGER} variable declared \kw{BETWEEN} $0$ \kw{AND} $\rho + 1$; PaQL's $\rho$ may be a constant or a data column, and here it is \texttt{max\_servings}, so each recipe is capped by its own value (a \querylang{} bound may likewise reference data columns, evaluated per candidate). \kw{REPEAT 0} is the keep-or-drop case and becomes \kw{SELECTION BINARY}, as in the cart of \Cref{sec:at-a-glance}; omitting \kw{REPEAT} leaves multiplicity unbounded and becomes a \kw{SELECTION INTEGER} variable with no upper bound. \kw{SELECTION} then supplies auto-multiply (\Cref{sec:package-semantics}), so aggregates count with multiplicity: \texttt{SUM(kcal)} reads as \texttt{SUM(kcal * servings)} and \texttt{COUNT(*)} as \texttt{SUM(servings)}. A recipe taken $k$ times contributes $k$-fold, matching PaQL's package semantics.

\querylang{} also expresses problems beyond package selection. A grouped constraint such as \texttt{SUM(weight) <= 20 BY category} (the \texttt{per\_category} constraint of \Cref{sec:example-knapsack}) has no PaQL form: \kw{SUCH THAT} states a single constraint over the whole package, whereas \kwBY{} (\Cref{sec:constraint-syntax}) generates one per group. Filtered constraints (\kwFILTER{}), companion decision variables with interval linking, multi-granularity joins (\Cref{sec:multi-granularity}), and continuous variables are likewise outside PaQL's scope.

\subsection{Limitations and Future Work}
\label{sec:future-extensions}

Three capabilities are outside the scope of this version. Each would extend the language with new constructs and new semantics, and we expect to add them as \querylang{} matures:

\begin{itemize}
\item \textbf{Multi-stage stochastic optimization.} Problems where decisions alternate with uncertain observations (e.g., decide inventory levels, observe demand, adjust orders) require a recourse or dynamic programming framework beyond the current single-stage model. Deterministic sequential decisions are already supported via decision chaining (\Cref{sec:composability}).
\item \textbf{Lazy constraint generation.} Some formulations have exponentially many constraints that a solver adds on demand during solving rather than stating up front, as with the subtour-elimination constraints of the traveling salesman problem~\cite{dfj-tsp} or with cutting planes. Generating constraints this way relies on solver callbacks the current language does not expose.
\item \textbf{Multi-objective optimization.} The current language supports a single objective. An extension such as \texttt{MINIMIZE (cost, -profit) PARETO} would return the Pareto frontier as a set of non-dominated solutions, each materialized as a separate result relation.
\end{itemize}


\clearpage
\bibliographystyle{plainnat}
\bibliography{references}

\end{document}